\documentclass[american]{aa}

\usepackage{adjustbox}

\usepackage{subcaption}

\usepackage{xcolor}

\usepackage{bm}

\usepackage{multirow}

\usepackage{placeins}

\usepackage{graphicx}
\usepackage{txfonts}
\usepackage[colorlinks=true,linkcolor=blue,citecolor=blue]{hyperref}

\begin{document}

   \title{X-ray polarization in radio-quiet AGN: Insights from the wedge corona model using Monte Carlo simulations}
   \titlerunning{Wedge corona model insights}

   \author{D. Tagliacozzo,
          \inst{1}*
          S. Bianchi,\inst{1}
          V. Gianolli,\inst{2}
          A. Gnarini,\inst{1}
          A. Marinucci,\inst{3}
          G. Matt,\inst{1}
          F. Ursini,\inst{1}
          W. Zhang\inst{4}}
    \authorrunning{D. Tagliacozzo et al.}

   \institute{Dipartimento di Matematica e Fisica, Universit\`a degli Studi Roma Tre, via della Vasca Navale 84, 00146 Roma, Italy   
         \and
         Dep. of Physics and Astronomy, Clemson University, Kinard Lab of Physics, 140 Delta Epsilon Ct, Clemson, SC 29634, USA
         \and
             Agenzia Spaziale Italiana, Via del Politecnico snc, 00133 Roma, Italy
             \and
             National Astronomical Observatories, Chinese Academy of Sciences, A20 Datun Road, Beijing 100101, China\\
\email{*daniele.tagliacozzo@uniroma3.it}
             }

   \date{}

   \abstract{In this study, we present novel calculations of X-ray polarization from radio-quiet and unobscured active galactic nuclei (AGNs) using the Monte Carlo code \textsc{monk}, which includes all general and special relativity effects. Our geometric model, referred to as the ``wedge corona'', features a homogeneous cloud of electrons characterized by an aspect ratio of h/r and a radius that extends down to the innermost stable circular orbit around the central black hole (BH). Adopting the physical parameters of the Seyfert galaxy NGC~4151 as a baseline, we investigated various geometric and physical configurations of the BH-corona-accretion disk (AD) system, such as the coronal opening angle, temperature, optical depth, BH spin, and the inner radius of the disk. 
   Finally, we compared our calculations with results from the Imaging X-ray Polarimetry Explorer (IXPE) for NGC~4151, the only radio-quiet and unobscured AGN with significant polarization detected by IXPE, to constrain the system's geometric parameters within the framework of the wedge corona model.}

   \keywords{Black hole physics -- Accretion, accretion discs -- Polarization -- Galaxies: active -- Galaxies: Seyfert -- Monte Carlo Radiative Transfer simulations}

   \maketitle
%
%-------------------------------------------------------------------

\section{Introduction}
\label{intro}
The study of X-ray emission in active galactic nuclei (AGNs) entered a new era after the launch, in December 2021, of the NASA-ASI Imaging X-ray Polarimetry Explorer (IXPE; \citealt{weiss_ixpe_2022JATIS...8b6002W}). In its working band ($2-8$ keV), the emission from unobscured radio-quiet AGNs is dominated by radiation from the hot electron corona (i.e., the primary continuum; \citealt{shapiro1976ApJ...204..187S, suny_tita_corona1980A&A....86..121S, haardt_maraschi_corona1991ApJ...380L..51H}) surrounding the BH-AD system in the center of these galaxies. This structure, in which optical-UV photons from the disk are thought to be Comptonized by hot electrons \citep{haardt1993ApJ...413..507H}, has been widely studied via spectroscopy in recent decades. Facilities such as BeppoSAX, Suzaku, INTEGRAL, and especially NuSTAR  were able to determine the coronal physical properties, i.e., the electron temperature ($kT_e$), found to range between tens and hundreds of keV, and its Thomson optical depth $\tau$ (e.g., \citealt{dadina2007A&A...461.1209D, dero2012MNRAS.420.2087D, mali2014ApJ...782L..25M, galaxies6020044, tortorefId0, middei2019A&A...630A.131M, Kamraj_2022}). Moreover, some constraints on coronal size and location have been derived using time-lag techniques (e.g., \citealt{uttarticle, fa2017AN....338..269F, cab2020MNRAS.498.3184C}). However, since spectroscopy alone does not distinguish between different geometric configurations (e.g., \citealt{tortorefId0, middei2019A&A...630A.131M}), constraining its morphology remains an open issue. Since inverse-Compton scattering produces polarized radiation that depends on the geometry of the disk-corona system (\citealt{Tamborra:2018ebb, wenda2019ApJ...875..148Z, ursini2022MNRAS.510.3674U}), X-ray polarimetry offers a promising tool. 

Many geometric models reflecting different coronal origins have been proposed to date. Among them are the ``spherical lamppost,'' the ``conical outflow,'' and the ``slab'' above the disk. Their polarization signatures have been calculated using several radiative transfer codes (e.g., \citealt{sh2010ApJ...712..908S, Tamborra:2018ebb, wenda2019ApJ...875..148Z}) and compared with data collected by IXPE during its first years of operation (\citealt{Marinucci_2022,Gianolli2023,taglia2023MNRAS.525.4735T,ingram2023MNRAS.525.5437I, giano2024A&A...691A..29G}). The spherical lamppost consists of a spherical source placed at a certain height on the spin axis of the BH. Due to its high degree of symmetry, this configuration is expected to produce very low polarization degrees ($\Pi = 0 - 2 \%$) and polarization angle ($\Psi$) perpendicular to the AD axis (\citealt{ursini2022MNRAS.510.3674U}). The conical outflow model is commonly associated with an aborted jet (\citealt{hen1997A&A...326...87H, conerefId0}). According to this model, even in radio-quiet AGN, a jet can be launched, but it fails because its velocity is smaller than the escape velocity (\citealt{conerefId0}). In this scenario, $\Pi$ is expected to reach higher values than in the spherical lamppost model (up to $6\%$), while $\Psi$ remains perpendicular to the disk axis (\citealt{ursini2022MNRAS.510.3674U}), which is a common feature of vertically extended models. Another widely used geometric model is the slab corona (\citealt{liang1979ApJ...231L.111L, haardt_maraschi_corona1991ApJ...380L..51H}). In this model, the hot medium is assumed to extend over the AD and can occur when magnetic loops extend well above the disk plane and release energy through reconnection (e.g., \citealt{belo2017ApJ...850..141B}). This configuration is expected to produce a relatively high $\Pi$ (up to $14\%$), with a polarization angle parallel to the AD axis (\citealt{sve1996ApJ...470..249P, ursini2022MNRAS.510.3674U}). However, these models pose issues with respect to their ability to reproduce the spectra of radio-quiet AGNs and / or their physical origin and energy sustainability (\citealt{poutanen2018}). For these reasons, other geometric models have been proposed. 
In this paper, we discuss the so-called ``wedge corona model'' and present results of detailed Monte Carlo simulations assuming different physical and geometric configurations. 

The paper is structured as follows: in Sect. \ref{wedge}, we describe the Wedge model; in Sect. \ref{montecarlo}, we show the results of the Monte Carlo simulations, discussing the dependencies on the various parameters; in Sect. \ref{mcg} we apply the results to the case of the Seyfert galaxy NGC~4151, which is the only radio-quiet, unobscured AGN for which IXPE has so far detected a clear polarization; finally, in Sec \ref{discussion}, we summarize the results.

\section{The wedge}
\label{wedge}

\subsection{Why the wedge corona?}
\label{why}
As discussed in Sect. \ref{intro}, many geometric models have been developed for the AGN X-ray corona. The goal in developing such models is to reproduce spectral, timing, and polarimetric data, along with providing a reliable and convincing physical motivation.

With regard to X-ray polarization, recent observations of radio-quiet unobscured AGNs performed by IXPE place constraints on the coronal geometric configuration. During its first years of operations, IXPE observed four radio-quiet unobscured AGNs: MCG-05-23-16 (twice, in May and November 2022; \citealt{Marinucci_2022, taglia2023MNRAS.525.4735T}), IC~4329A (in January 2023; \citealt{ingram2023MNRAS.525.5437I}), NGC~4151 (twice, in December 2022 and in May 2024; \citealt{Gianolli2023, giano2024A&A...691A..29G}) and NCG 2110 (in October 2024).\\ 
In NGC~4151 a significant detection was obtained in both IXPE observations ($\Pi = 4.5\% \pm 0.9\%$ and $\Psi = 81^{\circ} \pm 6^{\circ}$, from a combined analysis). If most of the polarized light is attributed to the primary emission from the X-ray corona, as suggested by the spectral fitting, its polarization fraction reaches $7.1\% \pm 1.2\%$ and the polarization angle is consistent with the radio jet, favoring radially extended coronal models.
For IC~4329A,  IXPE measured polarization at a 2.97$\sigma$ confidence level, with $\Pi = 3.3\% \pm 1.1\%$ and $\Psi = 78^{\circ} \pm 10^{\circ}$, just slightly below the 3$\sigma$ threshold typically required for detection. From the contour plot, the most probable value of the polarization angle is consistent with the jet position angle inferred from a September 2021 ALMA (Atacama Large Millimeter Array) observation, again favoring radially extended coronal models.
In the case of MCG-05-23-16, the combined May and November 2022 observations yielded only an upper limit $\Pi < 3.2\%$ (at $99\%$ confidence level) to the polarization degree, and therefore the polarization angle, at this confidence level, is unconstrained. However, from the contour plot between the degree and angle of polarization, an alignment between $\Psi$ and the position angle of the \textsc{[OIII]} emission (a proxy for the AD axis) has the highest probability, again hinting at a radially extended corona.
Finally, NGC~2110 has been recently observed and its spectro-polarimetric analysis is still ongoing.

In summary, IXPE results show that due to the alignment between the polarization angle of the $2 - 8$ keV coronal emission (i.e. primary continuum) and the AD symmetry axis, radially extended coronal models are favored over vertically extended ones, for which angles perpendicular to the disk axis are expected. 

From a spectroscopic point of view, \cite{stern1995ApJ...449L..13S} showed that, in AGNs, slab models sandwiching the disk generally produce spectra steeper than observed, owing to the substantial flux of reprocessed photons that cool the electrons in the corona. To overcome this and other problems, models have been proposed in which the standard AD is truncated at a radius larger than the innermost stable circular orbit (ISCO), and the X-ray-emitting corona takes the form of an inner hot, possibly magnetized accretion flow \citep{jedsadPetrucci_2008, poutanen2018}.

\subsection{The wedge corona model}
\label{model}
\begin{figure}[t]
\includegraphics[width=9cm] {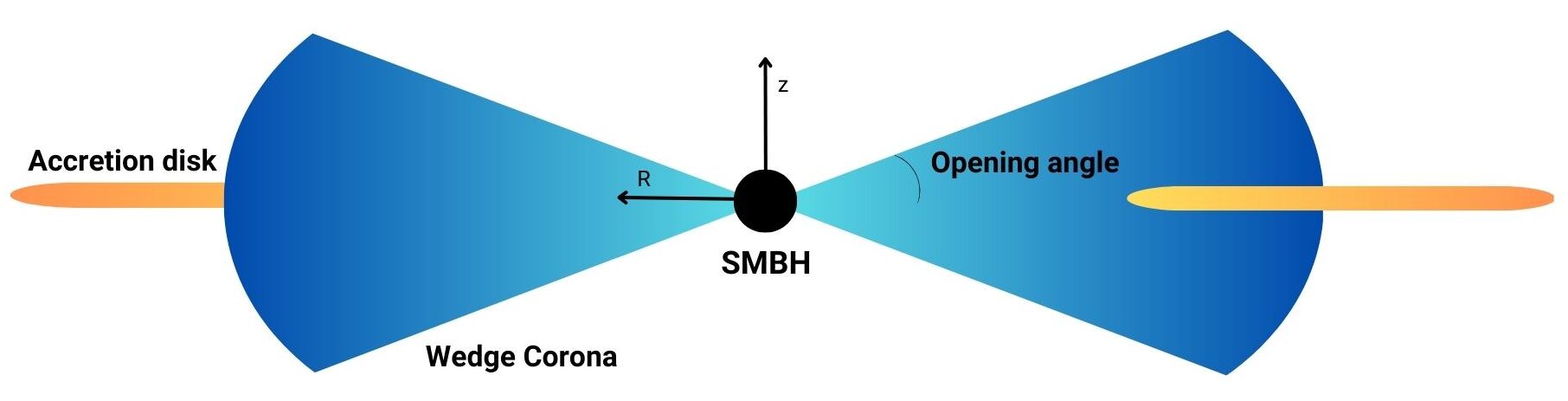}
\captionsetup{font=small}
\caption{Wedge corona model geometry, characterized by an inner ($R_{\textnormal{in}}$) and an outer ($R_{\textnormal{out}}$) radius and an opening angle ($\alpha$), measured from the AD plane. In the left configuration, the inner radius of the AD coincides with the outer radius of the corona, while in the right configuration it extends into the corona itself. The $R$ and $z$ axes represent the radial and the vertical coordinates, respectively.}\label{wedge_geometry}
\end{figure}

In this context, we describe the wedge corona model (\citealt{esin1997ApJ...489..865E, esin1998ApJ...505..854E, sh2010ApJ...712..908S, poutanen2018}; see Fig. \ref{wedge_geometry}). This model has the advantage of reducing the disk area subtended by the hot corona, thereby overcoming the problem of excessively steep spectra.  Moreover, this model results in polarization angles that are parallel to the AD vertical axis (\citealt{Gianolli2023, taglia2023MNRAS.525.4735T}; see Sect. \ref{montecarlo}), consistent with findings by IXPE in radio-quiet, unobscured AGNs.

The wedge geometry is defined by the following parameters: an inner radius ($R_{\textnormal{in}}$), an outer radius ($R_{\textnormal{out}}$), and a constant opening angle ($\alpha$) measured from the AD plane. We assume that the inner radius coincides with the ISCO, which depends on the spin value of BH, being equal to 6 $R_{\rm G}$ for $a=0$ and $\sim$1.24 $R_{\rm G}$ for $a=0.998$, where $R_{\rm G}$ is the gravitational radius. Unlike the slab configuration, the height ($h$) of the wedge increases with radius ($r$), maintaining a constant value of $\tan\alpha = h/r$. In this configuration, the AD is assumed to be truncated at a certain radius ($R_{\textnormal{disk}}$), while the corona represents a ``hot accretion flow'' extending down to the ISCO. A limitation of the present version of the \textsc{monk} code is that the radial density and temperature profiles of the wedge corona are assumed to be uniform. This may hamper the comparison between different BH spins. In fact, for high black hole spin, energy dissipation is expected to peak at lower radii, where the general relativity effects are stronger. If emission primarily originates from these inner regions, it could lead to substantial deviations in the polarization angle ($\Psi$). Additionally, different temperatures and densities close to the event horizon may significantly influence the expected polarization degree ($\Pi$). These effects should be considered in future investigations, together with a self-consistent treatment of the disc-corona interaction. The Thomson optical depth ($\tau$) is defined radially; it is computed along a straight line starting from the center of the system along the AD plane. The AD truncation radius can coincide with the external edge of the corona, be truncated at higher distances, or reach lower values down to the ISCO. Finally, the wedge can be either stationary (i.e., the angular velocity of the fluid is equal to that of a zero-angular momentum observer) or corotating with the Keplerian disk.

\section{Monte Carlo simulations}
\label{montecarlo}
To calculate the polarization properties of the X-ray primary continuum in the IXPE energy band (i.e., $2-8$ keV) as a function of the physical and geometric parameters of the wedge corona, we performed detailed Monte Carlo simulations with the radiative transfer code for Comptonized spectra \textsc{monk} \citep{wenda2019ApJ...875..148Z}. The \textsc{monk} code calculates the spectrum and polarization of Comptonized radiation from a corona illuminated by a standard AD. The seed optical and UV photons are generated according to the Novikov-Thorne disk emissivity, with a polarization parallel to the disk surface in the local frame, ranging from zero (face-on) to 11.7\% (edge-on) as expected for a semi-infinite, plane-parallel, scattering-dominated atmosphere \citep{1960ratr.book.....C}. After the photons leave the disk, they follow geodesics paths in Kerr spacetime around the black hole, either reaching the distant observer, being deflected back to the disk, or being lost by crossing the event horizon.  Photons that enter the corona may undergo Compton up-scattering, producing a Comptonized spectrum and modifying their polarization. The scatterings that occur in the corona significantly impact the radiation emitted by the disk in two key ways. Firstly, they modify the original spectrum through inverse Compton scattering, which increases the photon energy to higher values (mostly from optical and UV to X-ray), resulting in a harder spectrum. Secondly, they can alter both the intensity and alignment of the overall polarization signature at higher energies. 
The Stokes parameters of the scattered photons are calculated in the electron rest frame before being transformed to the observer's frame (Boyer-Lindquist). By counting the photons that reach the observer, a flux and polarization spectra are created. Since scattering produces linearly polarized photons, the code calculates the Stokes parameters $Q$ and $U$, while $V$ is assigned a value of zero.

Among the radio-quiet, unobscured AGNs observed by IXPE, NGC~4151 presents the most robust detection so far. Thus, this Seyfert galaxy was selected as the prototype source for the Monte Carlo simulations. NGC~4151 has been classified as a changing-look AGN \citep{penston84, puccetti07, shapovalova08}, going from $F_{\rm{0.5-10 \; keV}}$ $\sim$ 8.7$\times$ 10$^{-11}$ erg s$^{-1}$ cm$^{-2}$ in a low-flux state to $F_{\rm{0.5-10 \; keV}}$ $\sim$ 2.8$\times$ 10$^{-10}$ erg s$^{-1}$ cm$^{-2}$ in a high-flux state \citep{antonucci83, shapovalova12, Beuchert2017}. This source has been widely observed in X-rays and exhibits spectral variability and complex absorption structures (\citealt{Beuchert2017}). A near-maximal BH spin has been inferred from relativistic reflection off the AD \citep{cackett14, Keck2015, Beuchert2017}. With a BH mass of $M_{\rm BH}=4.57\times10^7M_{\odot}$, measured via optical and UV reverberation (\citealt{bentz06}), this source has an Eddington ratio of $1\%$ (\citealt{Keck2015}). From the joint observations of NGC~4151, performed with {\it XMM-Newton}, {\it NuSTAR}, and IXPE in December 2022, \citet{Gianolli2023} found, using \textsc{nthcomp} to model the Comptonized primary continuum, a spectral index $\Gamma=1.85\pm0.01$ and an electron temperature $kT_{\rm e}=60^{+7}_{-6}$ keV. As  explained in Sect. \ref{why}, spectro-polarimetric analysis yielded a significant detection in both IXPE observations (from a combined analysis: $\Pi = 4.5\% \pm 0.9\%$ and $\Psi = 81^{\circ} \pm 6^{\circ}$, aligned with the AD axis; \citealt{giano2024A&A...691A..29G}). Finally, reverberation studies of the broad-line region, suggest a disk inclination of $58.1^{\circ +8.4}_{-9.6}$ for NGC~4151 \citep{bentz2022ApJ...934..168B}.

We began constructing a baseline wedge corona model by adopting the BH mass and Eddington ratio of NGC~4151. The coronal internal radius was set to the ISCO (6 $R_{\rm G}$ for $a=0$) and the external radius to 25 $R_{\rm G}$. Initially, we set the AD inner radius to 25 $R_{\rm G}$, coinciding with the coronal external radius; the disk outer radius to 100 $R_{\rm G}$; and the wedge opening angle to the intermediate value of $30^{\circ}$. Additionally, we set the electron temperature to $60$ keV and determined the coronal optical depth such that it reproduced a primary continuum spectral index of $1.85$ ($\tau=1.9$), consistent with observations of NGC~4151. Both the disk and the corona rotate with Keplerian velocity around the central BH.\\

\subsection{Opening angle}
\label{opening_angles_comparison}
\begin{figure}[t]
\includegraphics[width=8.15cm] {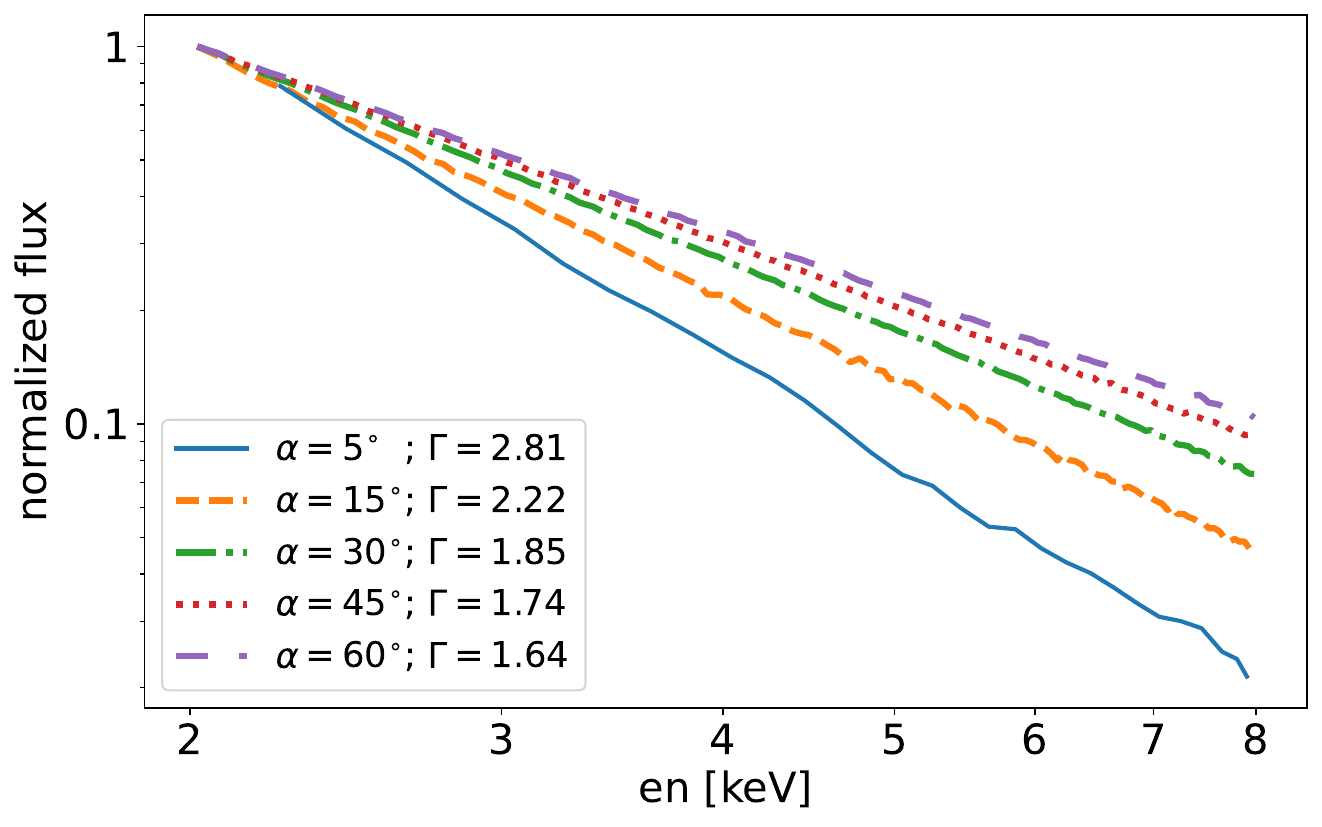}
\captionsetup{font=small} 
\caption{Primary continuum normalized flux as a function of  photons energy ($2-8$ keV) for different wedge opening angles ($\alpha$) in the baseline configuration. The parameters are: $R_{corona}=6 R_G$, $R_{disk}=25 R_G$, $kT_e=60$ keV, and $\tau=1.9$. As the opening angle increases, the spectral index ($\Gamma$) of the nuclear emission drops, leading to harder spectra.}\label{I_en_sk_a0_diff_ap_tau_const}
\end{figure}

Starting from the baseline configuration, we modified the wedge opening angle, testing four different cases: $5^{\circ}$, $15^{\circ}$, $30^{\circ}$, $45^{\circ}$, and $60^{\circ}$. We also tested different values of the BH spin ($a=0$ and $a=0.998$), modifying the coronal internal radius accordingly (6 $R_{\rm G}$ for $a=0$ and 1.24 $R_{\rm G}$ for $a=0.998$). For both spin cases, we find that the Comptonized spectrum hardens with increasing opening angle. This hardening occurs because increasing the opening angle increases the effective optical depths experienced by the soft photons, increasing the number of scatterings a photon undergoes before leaving the corona, and thus resulting in a hardening of the spectrum. Figure\ref{I_en_sk_a0_diff_ap_tau_const} shows the primary continuum spectral shape as a function of the energy in the $2-8$ keV energy band for different values of the wedge opening angle. 

\begin{figure}[h!]
\includegraphics[width=8.15cm] {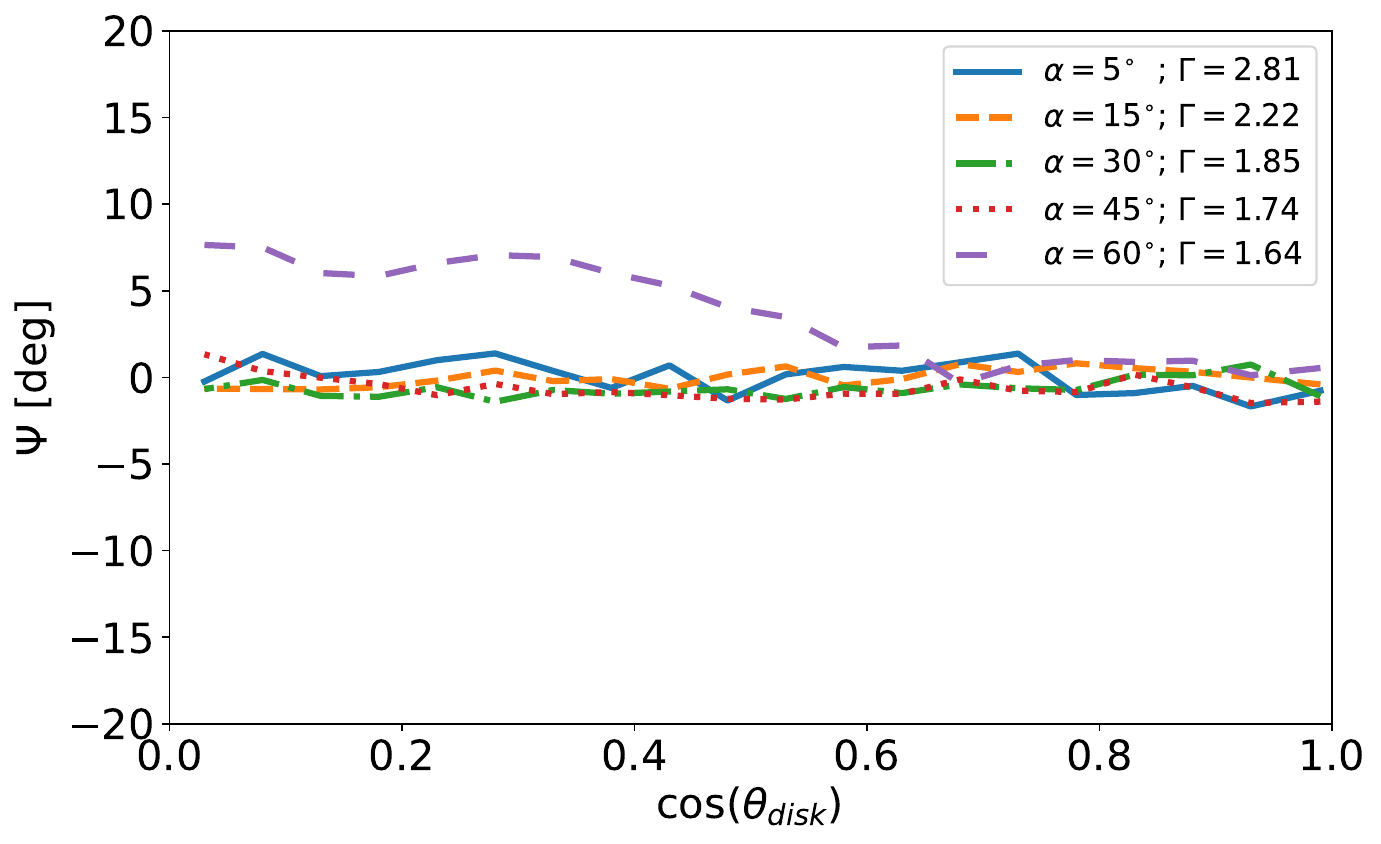}
\captionsetup{font=small}
\caption{Polarization angle ($\Psi$), summed between 2 and 8 keV, as a function of the inclination of the source ($\cos{\theta_{disk}}=1$ corresponds to the face-on view), assuming different wedge opening angles ($\alpha$) in the same configuration as in Fig. \ref{I_en_sk_a0_diff_ap_tau_const}. The polarization angle, $\Psi$, is generally parallel to the AD axis, except for $\alpha=60^{\circ}$ (dotted purple line).}\label{PA_opang}
\end{figure}

Regarding the polarization properties, we found that, consistent with  \cite{taglia2023MNRAS.525.4735T}, $\Psi$ is generally parallel to the AD axis (i.e., $\Psi=0$). Figure  \ref{PA_opang} shows $\Psi$, summed between 2 and 8 keV, as a function of the inclination angle (measured from the disk axis) for different wedge opening angles. We observe a slight deviation for $\alpha=60^{\circ}$, where, at high source inclinations ,$\Psi$ assumes values that increasingly differ from zero. This is because, locally, photons emerging from the corona have mean polarization vectors preferentially directed either perpendicular or parallel to the coronal surface. For high opening angles,  this direction differs significantly from that of the disk axis. In the absence of rotation, summing all these photons results in a net polarization angle equal to zero for symmetry reasons, as we verified by imposing zero rotation of the corona. However, when rotation is included, Doppler boosting implies that approaching matter appears brighter than receding matter, effectively breaking the axial symmetry.

\begin{figure*}[h!]
    \centering
    \begin{subfigure}[b]{0.45\textwidth}
        \centering
        \includegraphics[width=8.15cm]{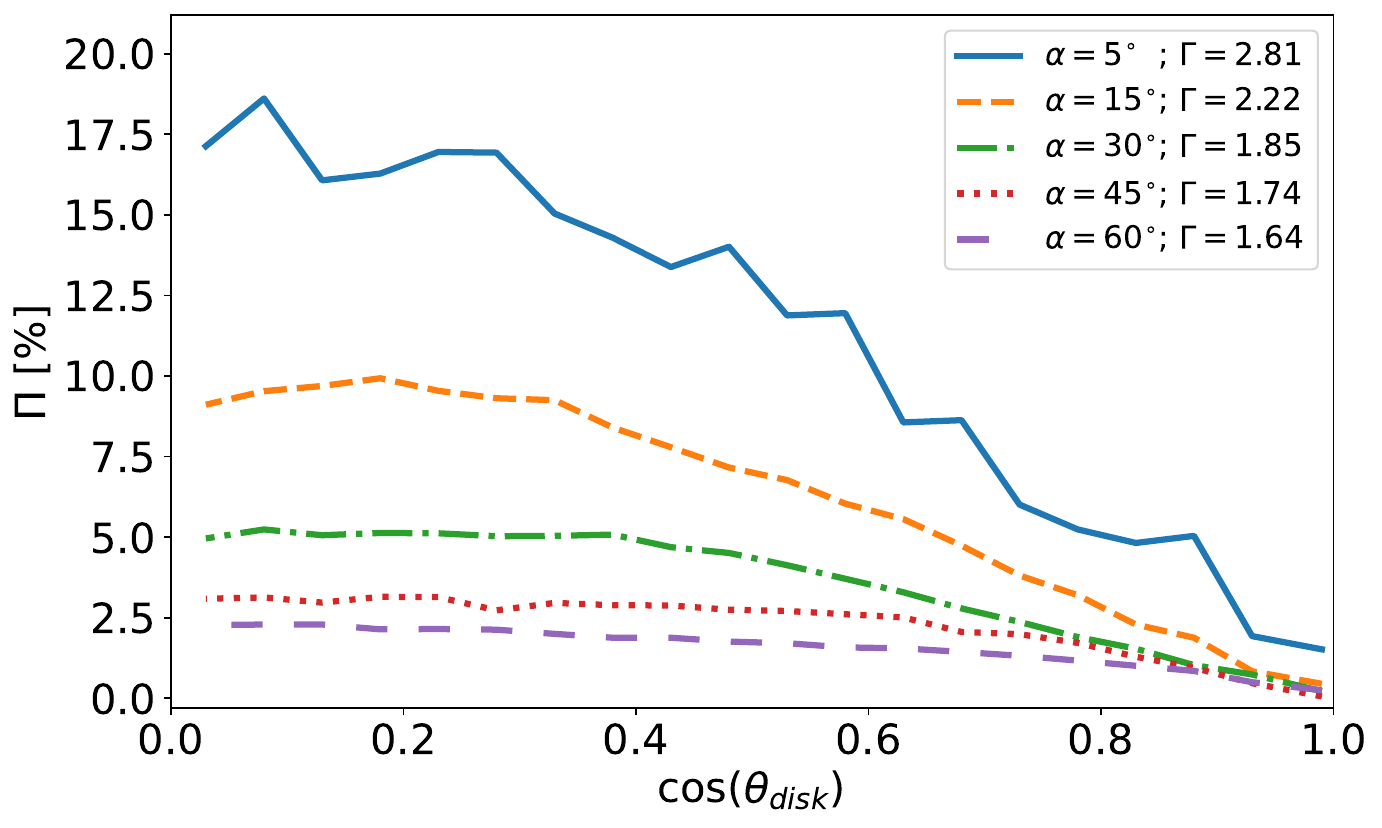} 
        \label{PD_sk_diff_ap_a0_tau_const}
    \end{subfigure}
    \hfill
    \begin{subfigure}[b]{0.45\textwidth}
        \centering
        \includegraphics[width=8.15cm]{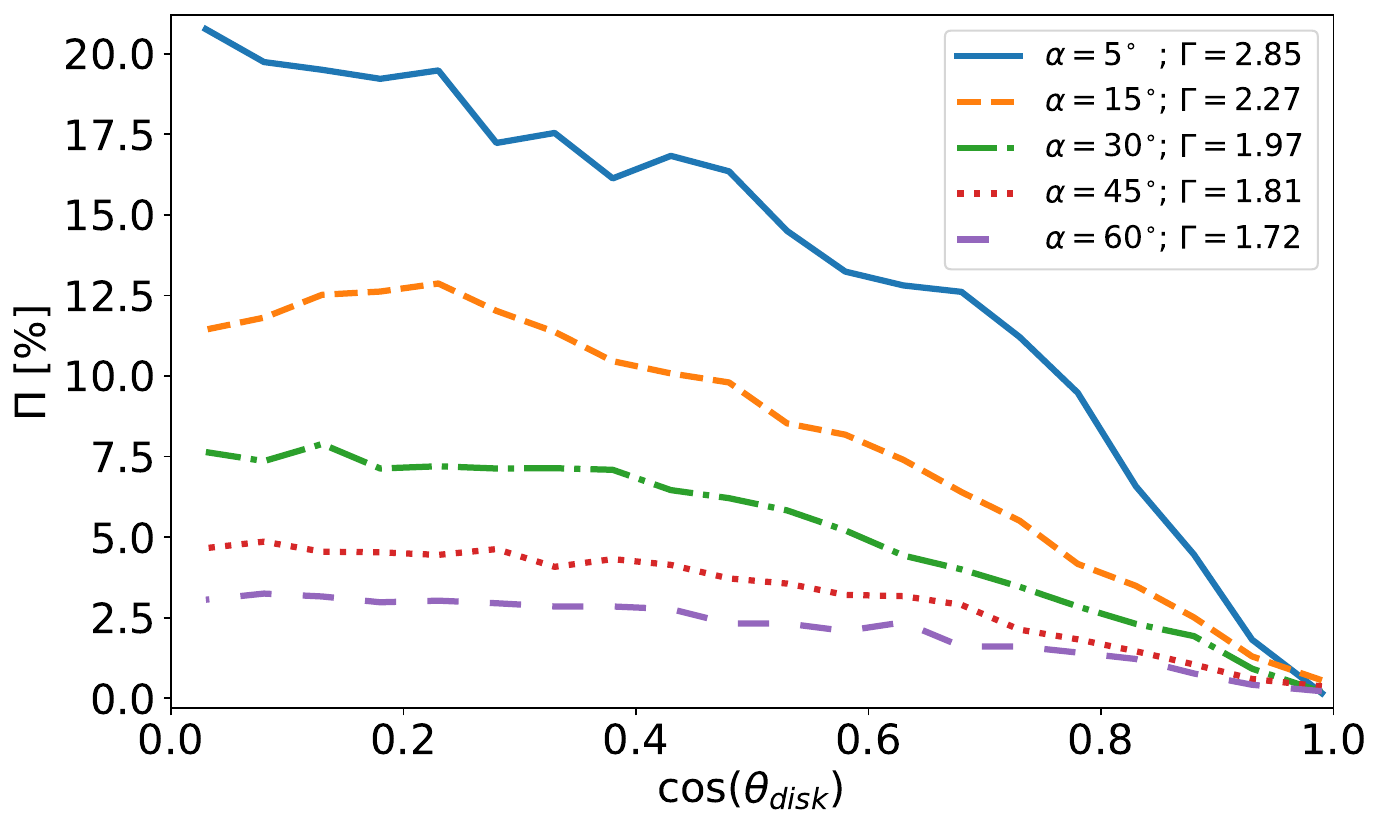} 
        \label{PD_sk_diff_opang_a998_tau_const}
    \end{subfigure}
 \captionsetup{font=small}
    \caption{Polarization fraction ($\Pi$), summed between 2 and 8 keV, as a function of the inclination of the source, assuming different wedge opening angles. In both panels, the coronal temperature is set to $kT_e=60$ keV and $R_{disk}=25 R_G$. The left panel corresponds to the $a=0$ and $R_{corona}=6 R_G$ scenario, while the right panel corresponds to $a=0.998$ and $R_{corona}=1.24 R_G$ scenario. As the opening angle increases, $\Pi$ decreases due to the increasing degree of symmetry of the system.}
    \label{diff_opang}
\end{figure*}

Regarding the polarization degree, $\Pi$ decreases with increasing opening angle owing to the rise in spherical symmetry, moving from a ``slab-like'' ($\alpha=5^{\circ}$) to a ``sphere-like'' ($\alpha=60^{\circ}$) configuration. Figure\ref{diff_opang} shows $\Pi$, summed between 2 and 8 keV, as a function of the inclination angle for different wedge opening angles, for both $a=0$ (left panel) and $a=0.998$ (right panel). The dependence on the BH spin is discussed in Sect. \ref{spin_comparison}.

\subsection{Disk inner radius}
\label{disc_sinking_comparison}
\begin{figure*}[h!]
    \centering
    \begin{subfigure}[b]{0.45\textwidth}
        \centering
        \includegraphics[width=8.15cm]{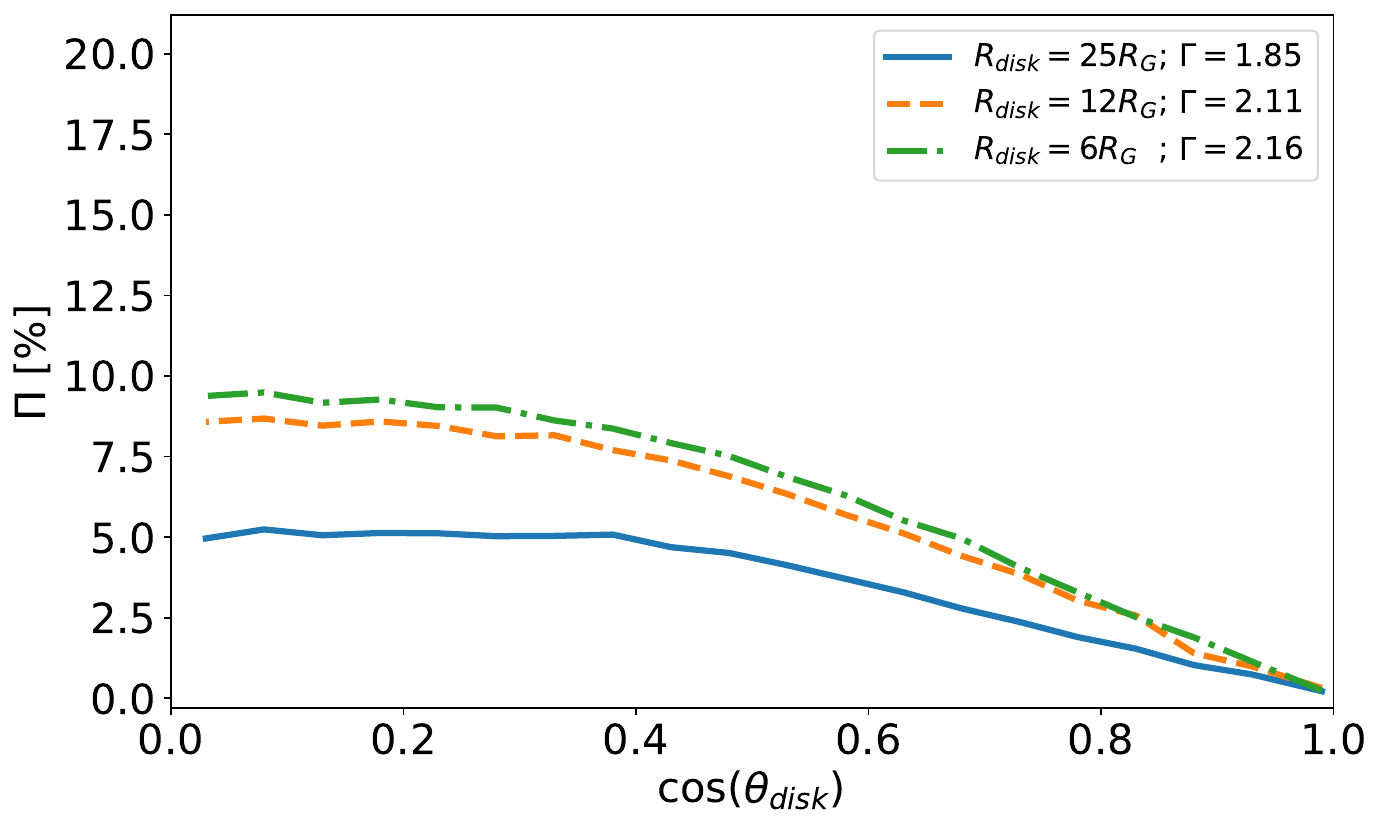} 
        \label{PD_sk_diff_disc_a0}
    \end{subfigure}
    \hfill
    \begin{subfigure}[b]{0.45\textwidth}
        \centering
        \includegraphics[width=8.15cm]{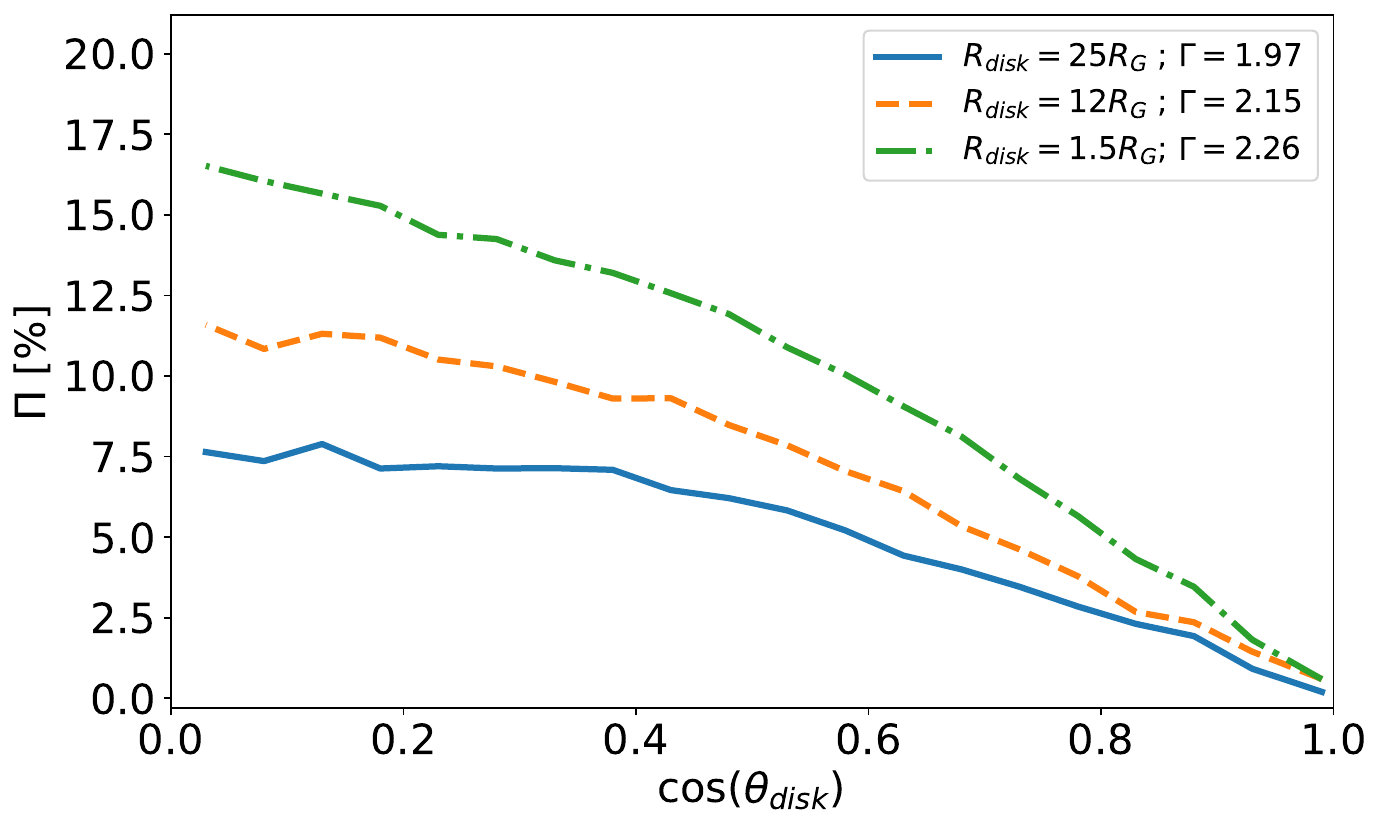} 
        \label{PD_sk_diff_disc_a998}
    \end{subfigure}
    \captionsetup{font=small}
    \caption{As in Fig. \ref{diff_opang}, but assuming different values for the AD internal radius ($R_{\textnormal{disk}}$) in the baseline configuration. In both panels, the coronal temperature is set to $kT_e=60$ keV and the wedge opening angle to $\alpha=30^{\circ}$. The left panel shows the $a=0$ and $R_{corona}=6 R_G$ scenario. The difference between $R_{\textnormal{disk}}=12R_{\rm G}$ and $R_{\textnormal{disk}}=6R_{\rm G}$ is small. The right panel shows the $a=0.998$ and $R_{corona}=1.24 R_G$ scenario. Here, the difference between $R_{\textnormal{disk}}=12R_{\rm G}$ and $R_{\textnormal{disk}}=1.24R_{\rm G}$ is much higher.}
    \label{diff_disc}
\end{figure*}

\begin{figure}
\includegraphics[width=8.15cm] {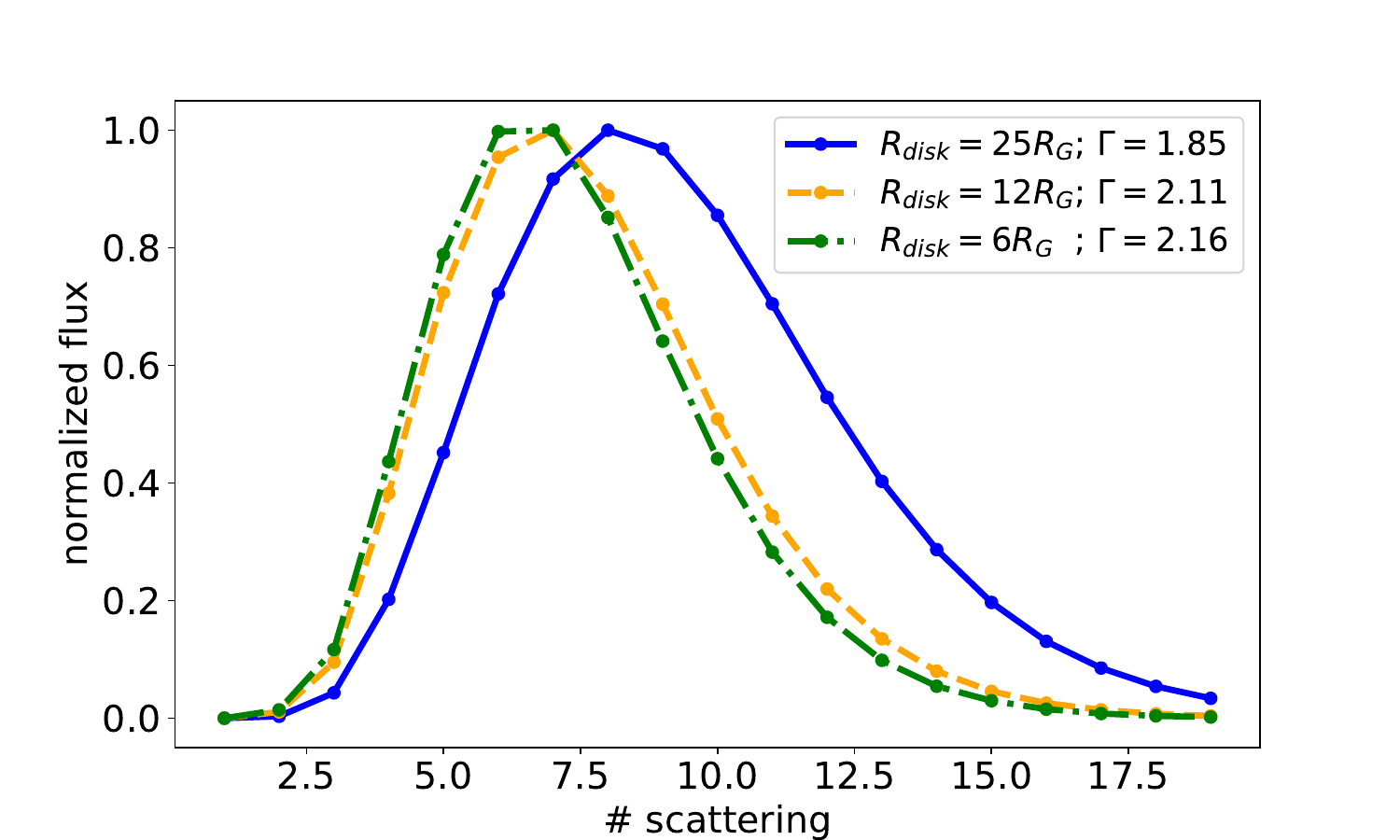}
\captionsetup{font=small}
\caption{Normalized flux as a function of photon scattering number, summed over the $2-8$ keV energy band. The coronal parameters are the same as in Fig \ref{diff_disc} (left panel). As shown, the dominant scattering level decreases for lower values of $R_{\textnormal{disk}}$.}\label{sca}
\end{figure}

Next, we tested different values of the inner radius of the AD ($R_{\textnormal{disk}}$). While in the baseline model $R_{\textnormal{disk}}$ only reaches the outer radius of the corona ($25$ $R_{\rm G}$), here we added two additional scenarios: in one, we set $R_{\textnormal{disk}}$ to approximately half of the wedge radius ($12$ $R_{\rm G}$), while in the other, we set it close to the ISCO ($6$ $R_{\rm G}$ for $a=0$ and $1.24$ $R_{\rm G}$ for $a=0.998$). We observed a softening of the primary continuum spectrum as $R_{\textnormal{disk}}$ goes deeper into the corona. This is due to the peculiar shape of the wedge, which causes the effective optical depth lessens when seed photons are emitted at low radii. For the same reason, $\Pi$ increases when $R_{\textnormal{disk}}$ decreases (see Fig. \ref{diff_disc}). This is illustrated in Fig. \ref{sca}, where the normalized flux, summed over the $2-8$ keV band, is plotted as a function of the number of scatterings undergone by photons inside the corona before escaping the system and reaching the observer. We note that, when the disk sinks into the corona, the dominant scattering number decreases. This is due both to the reduced effective optical depth mentioned above and to the hotter inner regions, which require fewer scatterings to shift the photon energy into the $2-8$ keV band. 

\subsection{Black hole spin}
\label{spin_comparison}
\begin{figure*}[h!]
    \centering
    \begin{subfigure}[b]{0.45\textwidth}
        \centering
        \includegraphics[width=8.15cm]{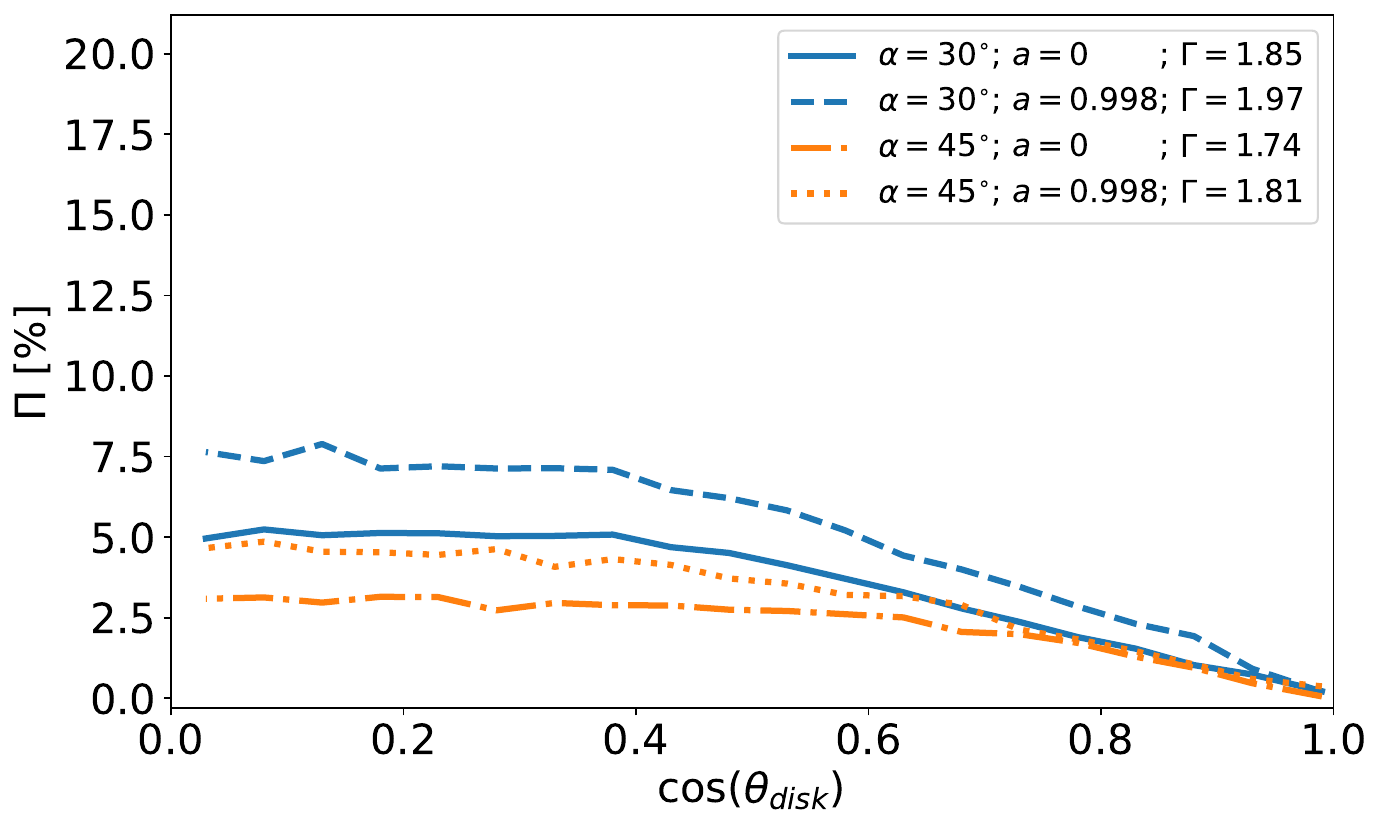} 
        \label{PD_sk_diff_spin_ext_30_45_deg}
    \end{subfigure}
    \hfill
    \begin{subfigure}[b]{0.45\textwidth}
        \centering
        \includegraphics[width=8.15cm]{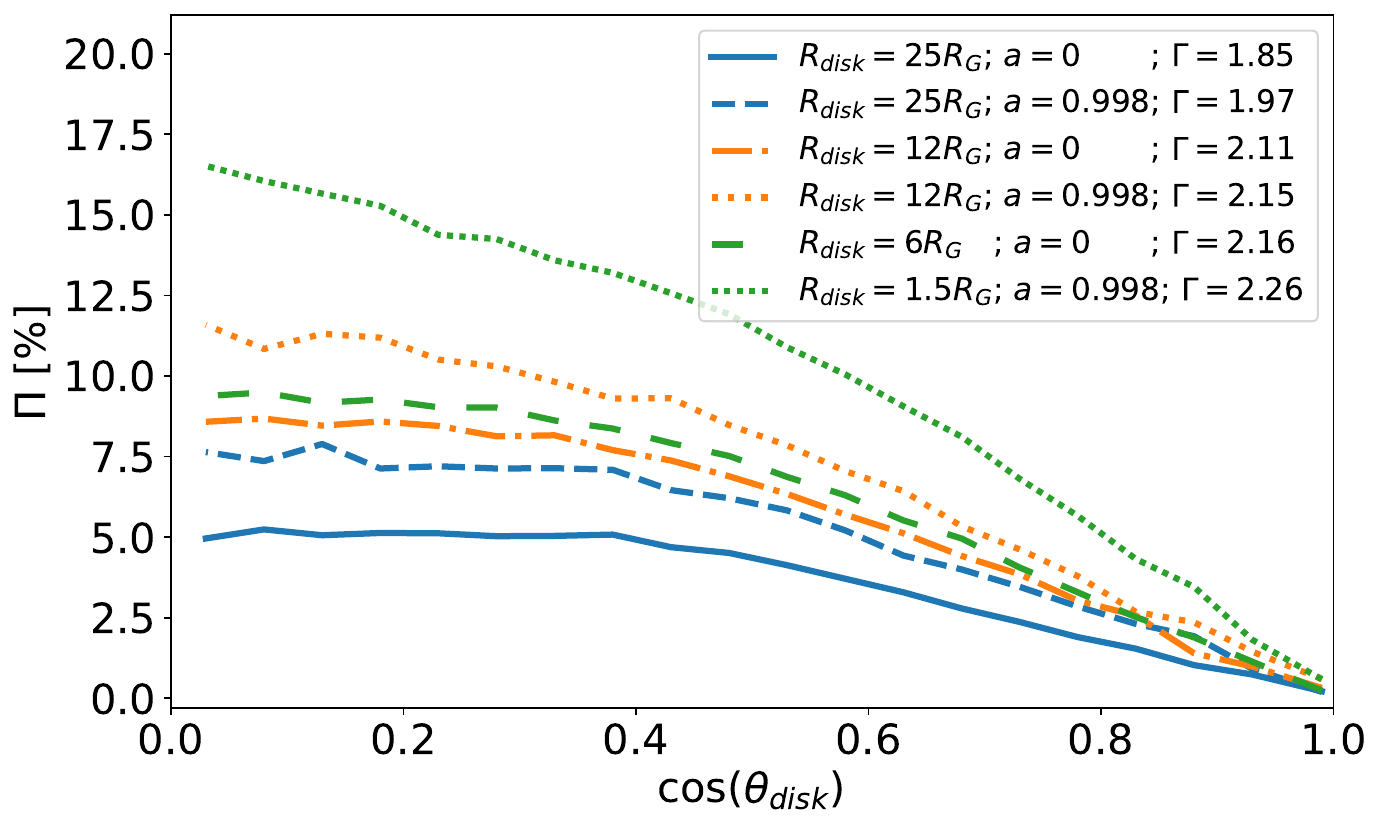} \label{PD_sk_diff_spin_int1_int2_ext30deg_deg}
    \end{subfigure}
    \captionsetup{font=small}
    \caption{As in Fig. \ref{diff_opang}, 
    but assuming different values for BH spin. In both panels, the coronal temperature is set to $kT_e=60$ keV and the inner radius of the corona is set to the ISCO value of the corresponding spin configuration. The left panel compares BH spins with $R_{disk}=25 R_G$ for different wedge opening angles ($\alpha$). The right panel compares BH spins for different values of the inner radius of the AD ($R_{disk}$). The $\Pi$ increases with increasing spin.}
    \label{diff_spin}
\end{figure*}

We also tested different values of the BH spin, exploring both $a=0$ and $a=0.998$. The inner radius of the corona was modified accordingly ($6$ $R_{\rm G}$ for $a=0$ and $=1.24$ $R_{\rm G}$ for $a=0.998$). The same modification was applied to the internal radius of the disk when it extends down to the ISCO. We observe a softening of the primary continuum spectrum and an increase in the polarization fraction for the high-spin cases. This is because, as shown by \cite{shak1973A&A....24..337S} and \cite{1973blho.conf..343N}, the disk temperature is higher at a given radius when the spin is higher.  Consequently, fewer scatterings are needed for the seed photons to be shifted into the X-ray band (see also \ref{disc_sinking_comparison}). Furthermore, the disk Keplerian velocity ($\Omega$) depends on the value of the BH spin according to $\Omega \sim (r^{3/2}+a)^{-1}$ \citep{wenda2019ApJ...875..148Z}. This determines an increasing loss of symmetry and a corresponding increase in the net polarization. The difference between $a=0$ and $a=0.998$ cases is particularly pronounced when the disk inner radius approaches values close to the ISCO for the same reason. Figure \ref{diff_spin} shows $\Pi$ as a function of the source inclination. The left panel compares the two BH spin scenarios in the baseline configuration (for $\alpha=30^{\circ}$ and $45^{\circ}$), while the right panel compares the two BH spins for different values of the disk inner radius.

\subsection{Coronal temperature}
\label{temperature_comparison}

In addition to the baseline case ($kT_{e}=60$ keV), we tested six more cases of coronal electron temperature: $kT_{e}= 25$, $75$, $100$, $200$, $300$, and $400$ keV. For all of these temperatures, we tested both the $a=0$ and the $a=0.998$ scenarios. We find that the spectrum was harder for higher values of the electron temperature \citep{middei2019A&A...630A.131M}, qualitatively following the well-known relation between $\Gamma$ and $kT_e$ for a fixed $\tau$ \citep{shapiro1976ApJ...204..187S, suny_tita_corona1980A&A....86..121S, ligh1987ApJ...319..643L}:
\begin{equation}
 \Gamma = [\frac{9}{4}+\frac{m_ec^2}{kT_e\tau(1+\tau/3)}]^{1/2}-1/2    .
\end{equation}

\begin{figure*}[h!]
    \centering
    \begin{subfigure}[b]{0.45\textwidth}
        \centering
        \includegraphics[width=8.15cm]{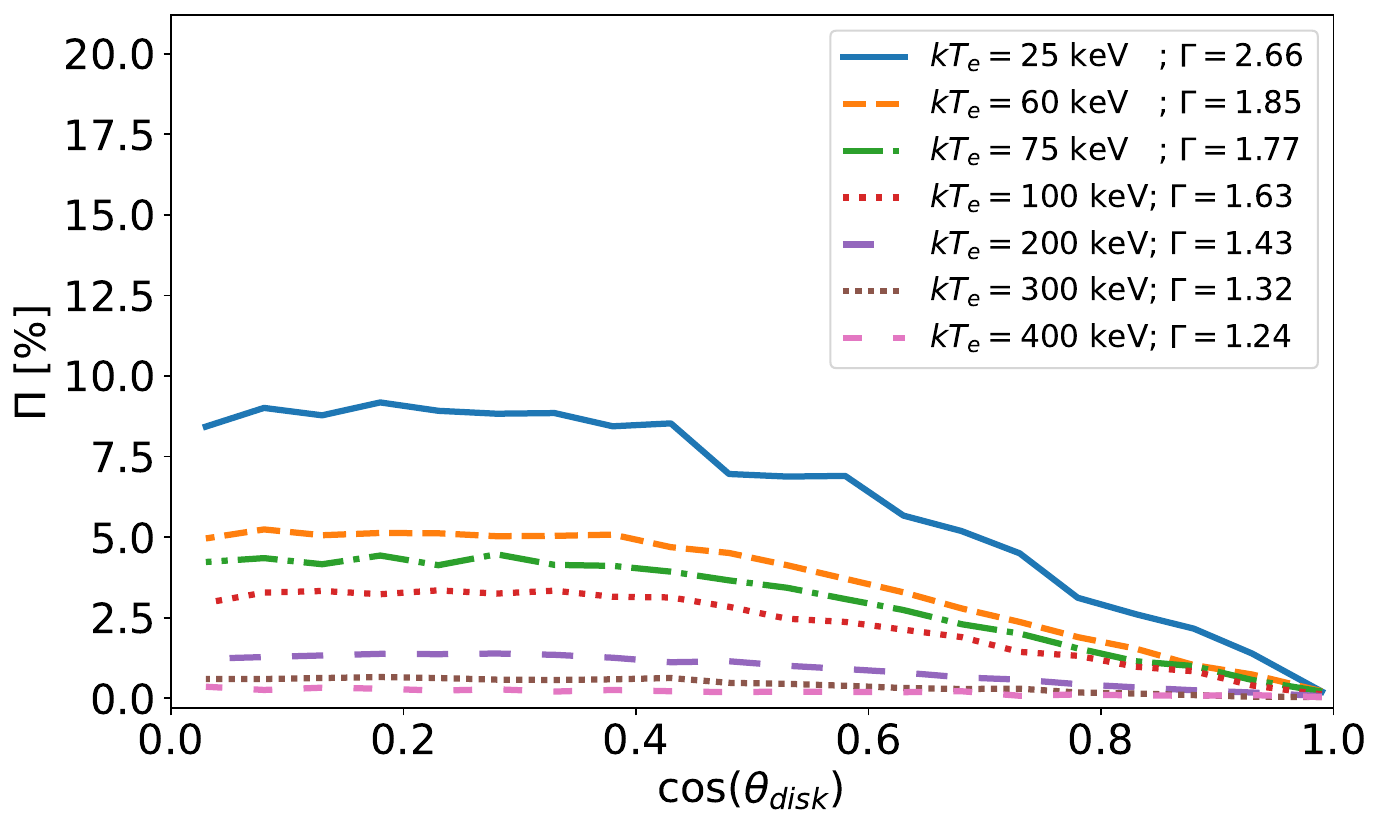} 
        \label{PD_sk_diff_kT_a0}
    \end{subfigure}
    \hfill
    \begin{subfigure}[b]{0.45\textwidth}
        \centering
        \includegraphics[width=8.15cm]{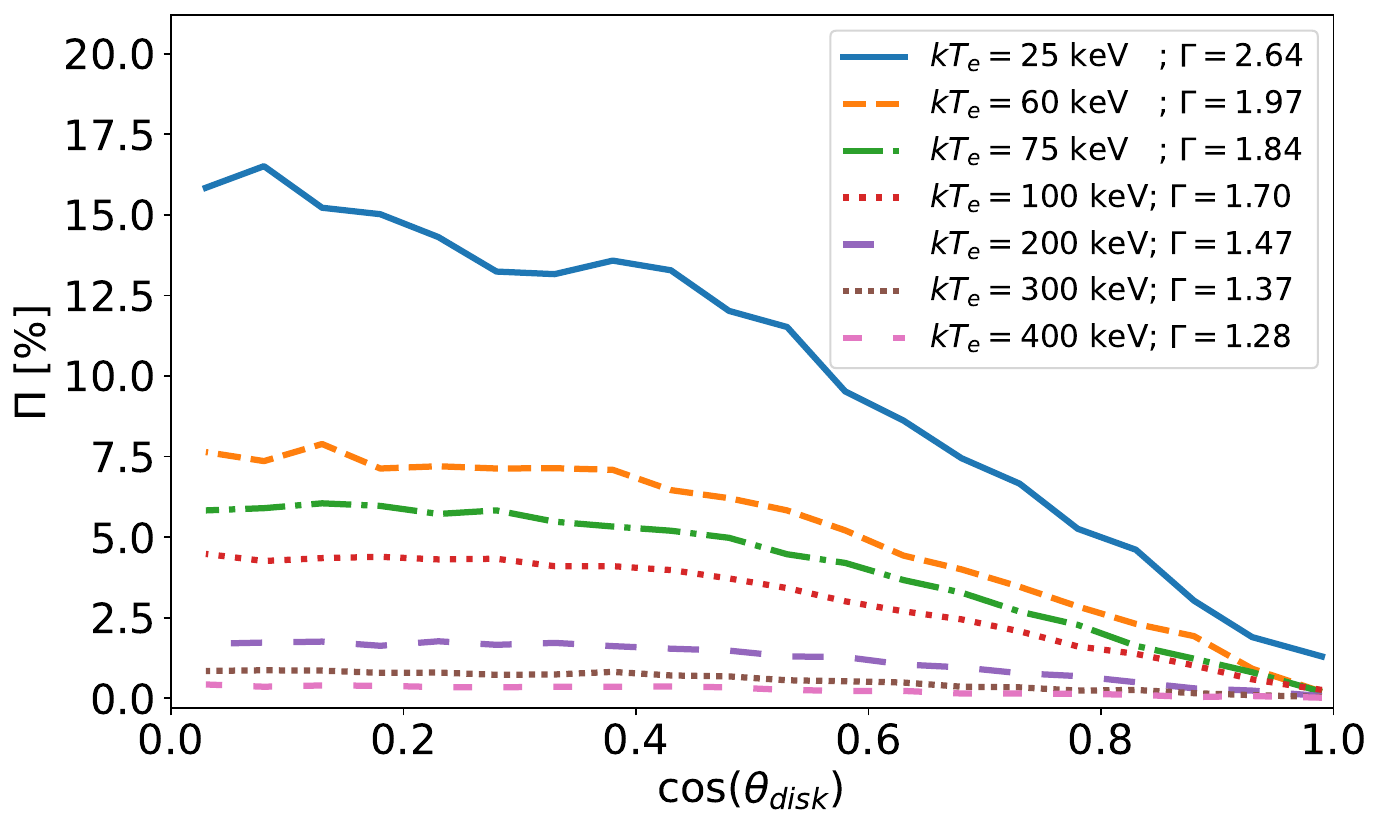} 
        \label{PD_sk_diff_kT_a998}
    \end{subfigure}
    \captionsetup{font=small}
    \caption{As in Fig. \ref{diff_opang}, but assuming different values for the coronal temperature. In both panels, the wedge opening angle is set to $\alpha=30^{\circ}$, the inner radius of the corona is set to the ISCO value of the corresponding spin configuration, and the inner radius of the AD is fixed at $R_{disk}=25 R_G$. The left panel shows the $a=0$ and $R_{corona}=6 R_G$ scenario. The right panel shows the: $a=0.998$ and $R_{corona}=1.24 R_G$ scenario . When the temperature increases, $\Pi$ decreases due to electrons motion randomization.}
    \label{diff_kT}
\end{figure*}

Regarding polarization, we observe a decrease in $\Pi$ moving from $kT_{e}=25$ keV to $kT_{e}=400$ keV. This is due to the well-known decrease in polarization with increasing electron energy in the Compton cross section \citep{poutemp1994ApJS...92..607P, MATT1996403}. Figure \ref{diff_kT}, presents $\Pi$ as a function of the source inclination, comparing different coronal temperatures ($a=0$ in the left panel and $a=0.998$ in the right).

\subsection{Optical depth}
\label{tau_comparison}
\begin{figure}
\includegraphics[width=8.15cm] {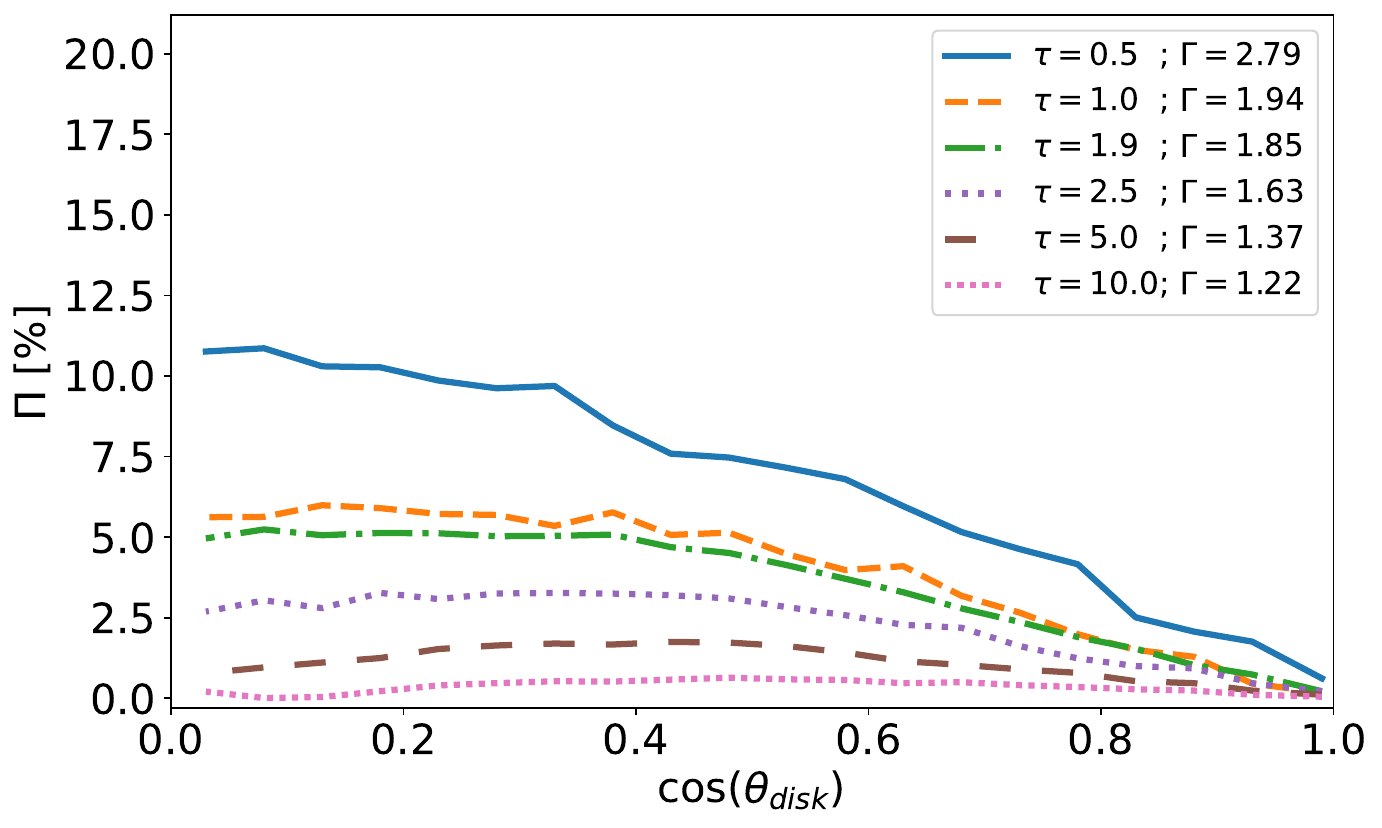}
\captionsetup{font=small}
\caption{As in Fig. \ref{diff_opang}, but assuming different values for coronal optical depth ($\tau$). In both panels, the wedge opening angle is set to $\alpha=30^{\circ}$, the spin of the BH is set to $a=0$, the inner radius of the corona is set to $R_{corona}=6 R_G$, and the inner radius of the AD to $R_{disk}=25 R_G$. When $\tau$ decreases, $\Pi$ increases.}\label{diff_tau}
\end{figure}

As a final test, we modified the baseline configuration by varying different coronal optical depth ($\tau$) values. We assumed $\tau=0.5$, $1$, $1.9$ (corresponding to the value reproducing the spectral index found for NGC~4151 in the baseline model), $2.5$, $5$, and $10$. As shown in Fig. \ref{diff_tau}, when $\tau$ decreases, $\Pi$ increases. As previously mentioned, this results from the fact that a high number of scatterings more effectively randomize the photon distribution.

\section{The case of NGC~4151}
\label{mcg}
\begin{figure*}[h!]
    \centering
    \begin{subfigure}[b]{0.45\textwidth}
        \centering
        \includegraphics[width=8.15cm]{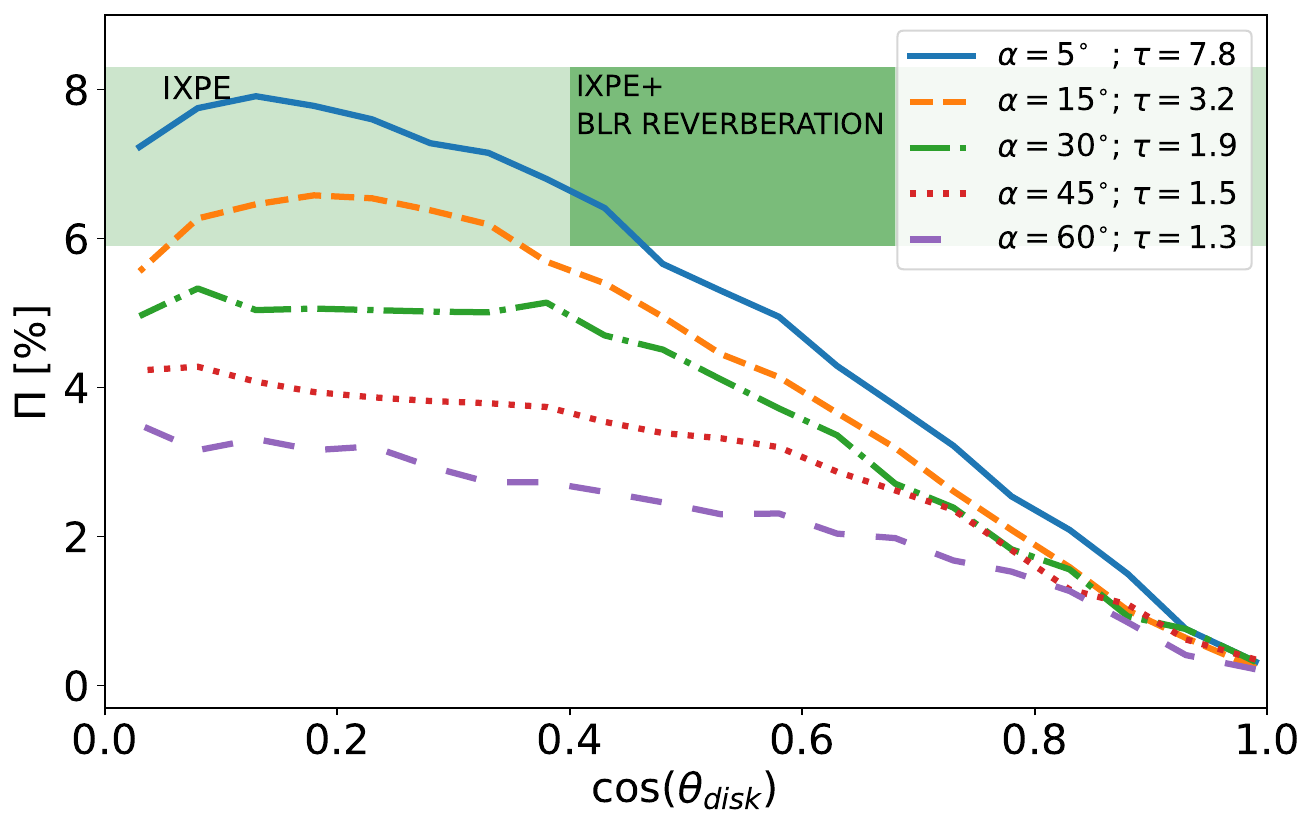} 
        \label{PD_sk_diff_ap_a0_gamma_const}
    \end{subfigure}
    \hfill
    \begin{subfigure}[b]{0.45\textwidth}
        \centering
        \includegraphics[width=8.15cm]{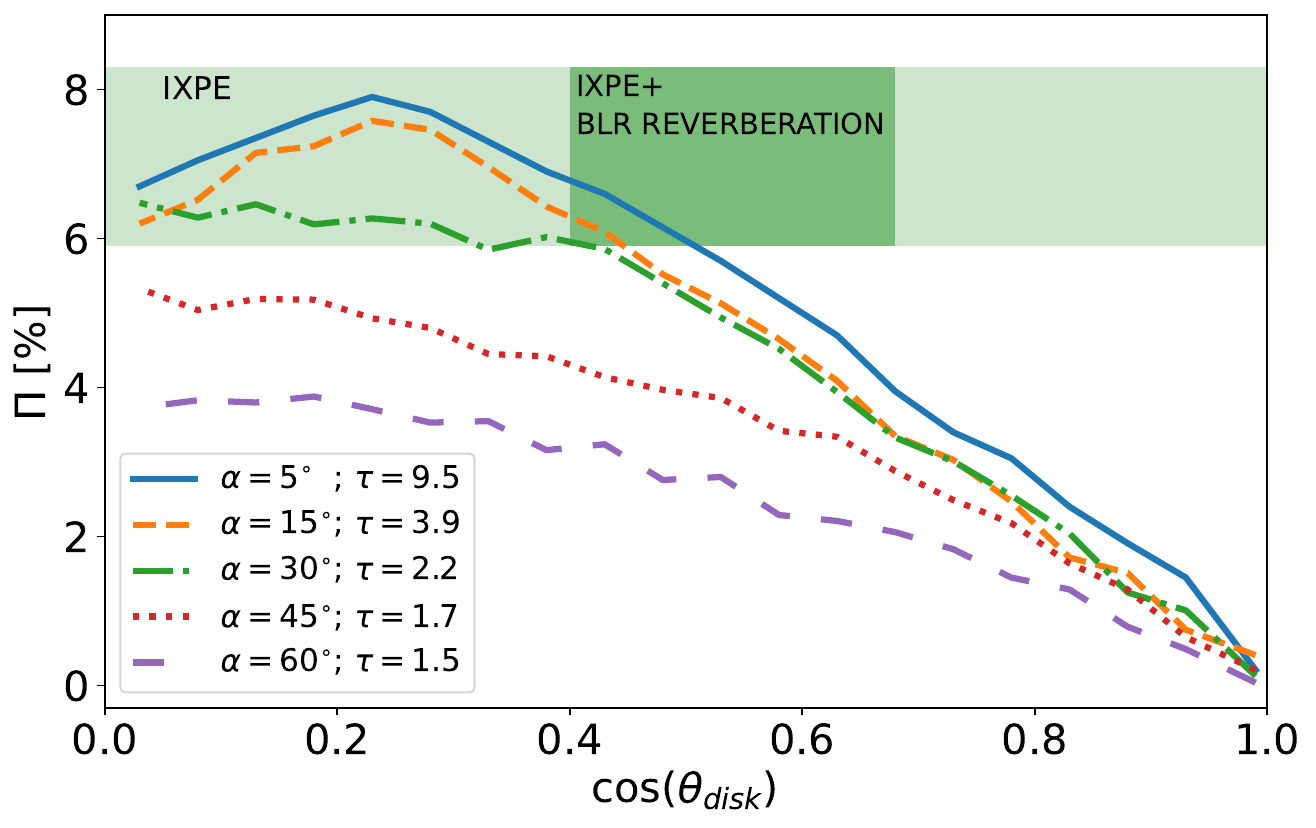} 
        \label{PD_sk_diff_opang_a998_gamma_const}
    \end{subfigure}
    \captionsetup{font=small}
    \caption{As in Fig. \ref{diff_opang}, but for a fixed spectral image value of $\Gamma=1.85$. The trend is similar to that found previously, although the differences are now dampened. The light green area represents the constraint from IXPE observations of NGC~4151, while the dark green area indicates the constraint resulting from the combination of IXPE observations and BLR reverberation. Both the constraints are given at the $68\%$ confidence level.}
    \label{diff_opang_gamma}
\end{figure*}

\begin{figure*}[h!]
    \centering
    \begin{subfigure}[b]{0.45\textwidth}
        \centering
        \includegraphics[width=8.15cm]{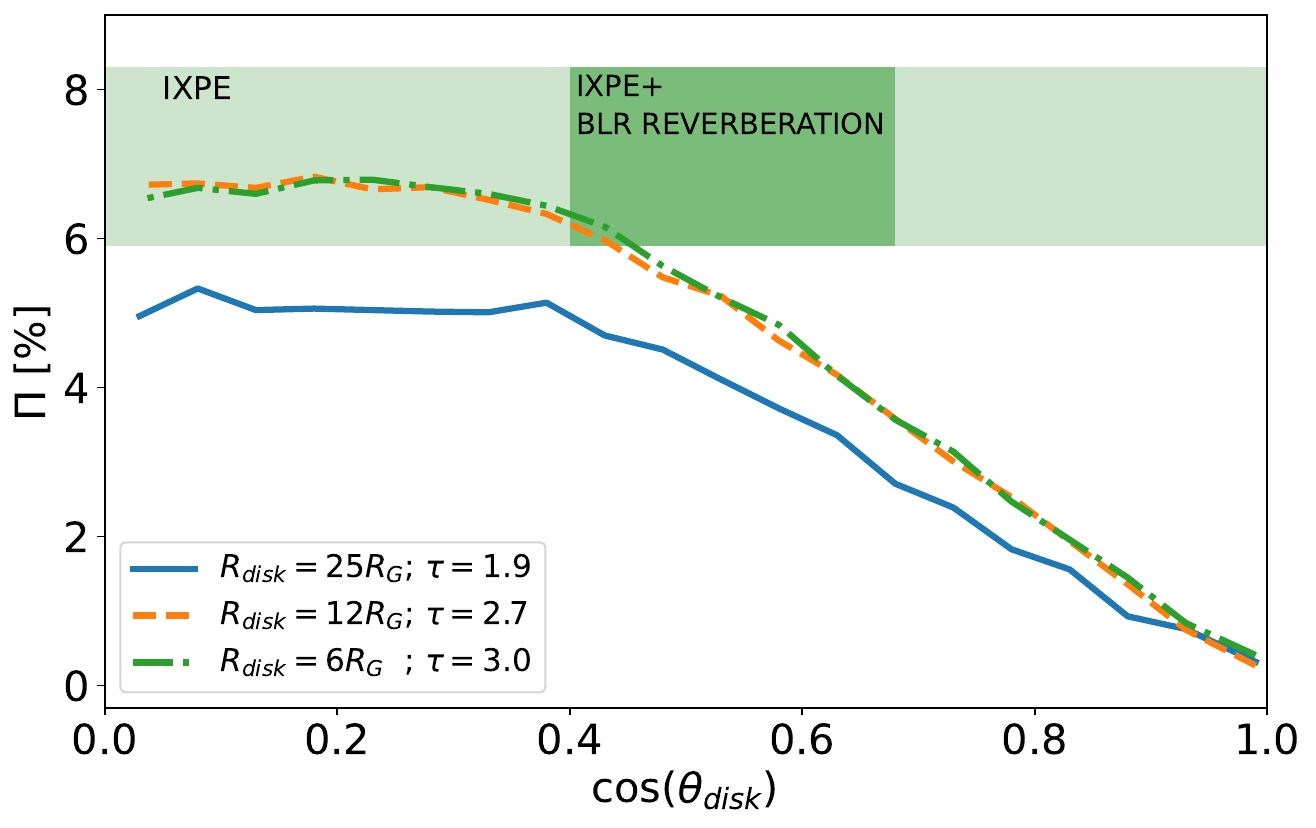} 
        \label{PD_sk_diff_disc_a0_gamma}
    \end{subfigure}
    \hfill
    \begin{subfigure}[b]{0.45\textwidth}
        \centering
        \includegraphics[width=8.15cm]{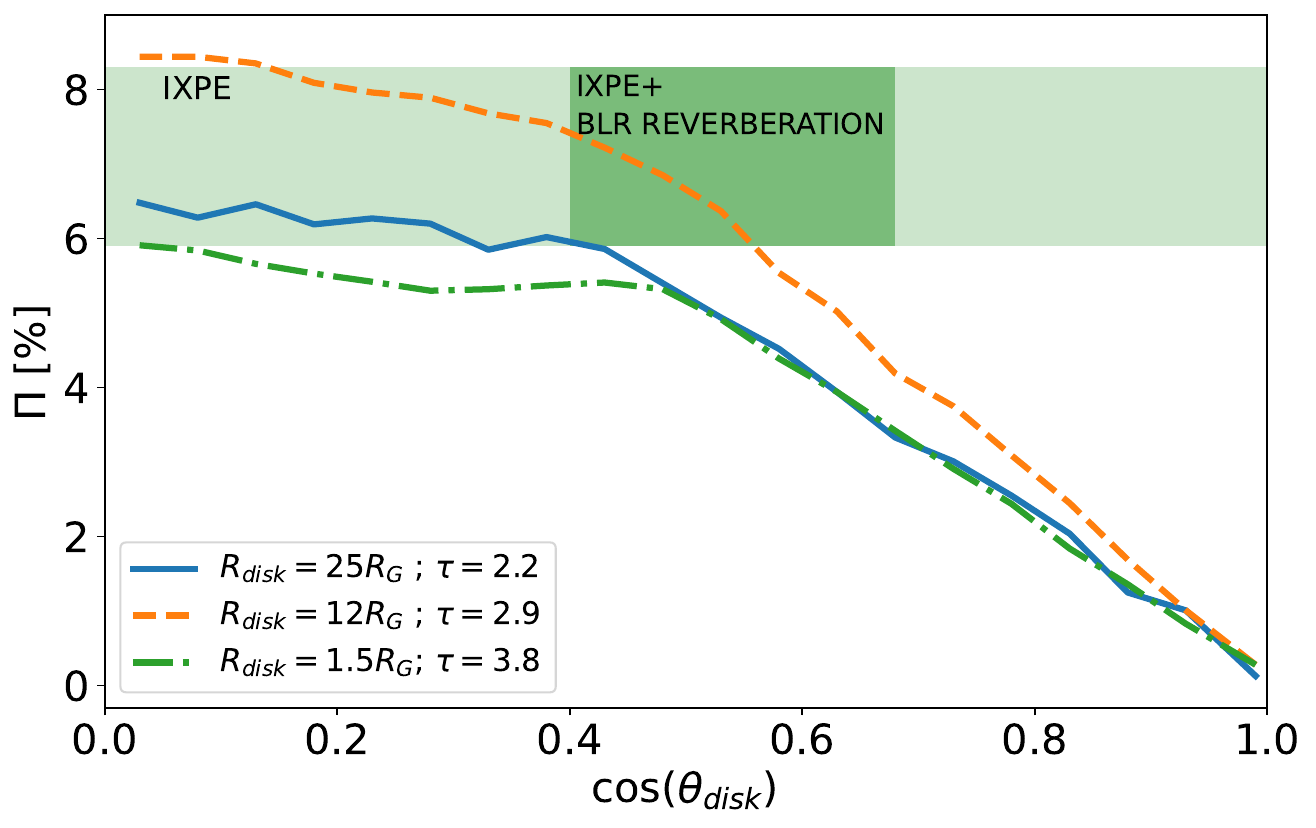} 
        \label{PD_sk_diff_disc_a998_gamma}
    \end{subfigure}
    \captionsetup{font=small}
    \caption{As in Fig. \ref{diff_disc}, but for a fixed spectral image value of $\Gamma=1.85$. The trend is similar, although the differences are now dampened.}
    \label{diff_disc_gamma}
\end{figure*}

\begin{figure*}[h!]
    \centering
    \begin{subfigure}[b]{0.45\textwidth}
        \centering
        \includegraphics[width=8.15cm]{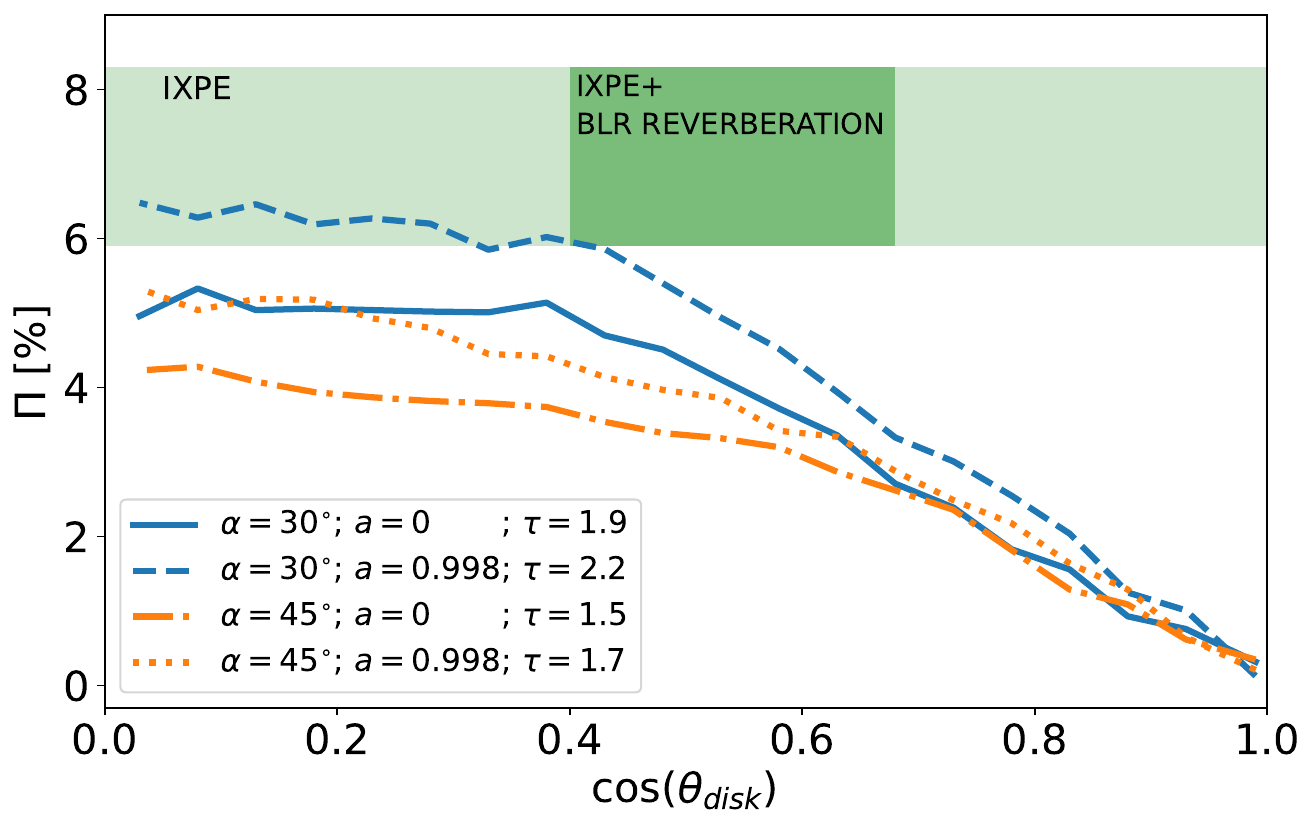} 
        \label{PD_sk_diff_spin_ext_30_45_deg_gamma}
    \end{subfigure}
    \hfill
    \begin{subfigure}[b]{0.45\textwidth}
        \centering
        \includegraphics[width=8.15cm]{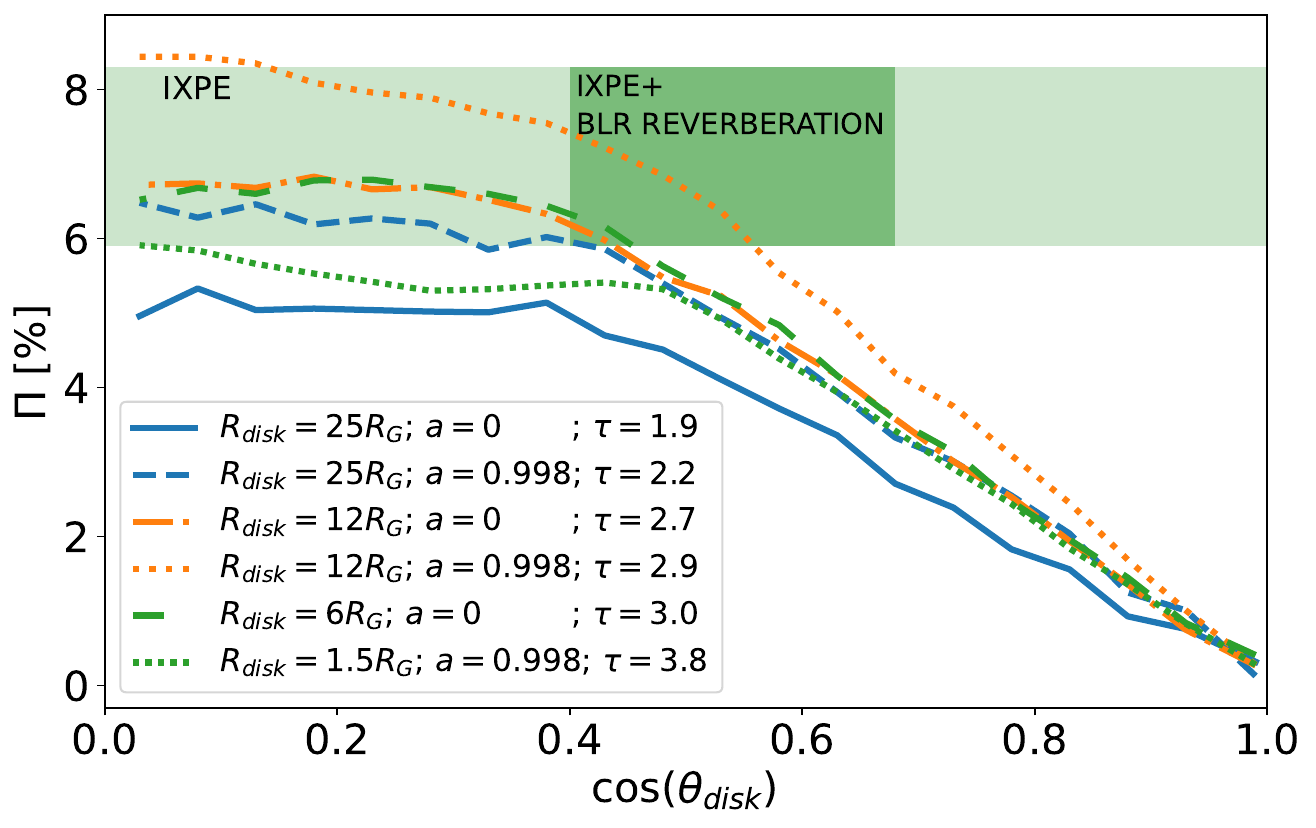} 
        \label{PD_sk_diff_spin_int1_int2_ext30deg_deg_gamma}
    \end{subfigure}
    \captionsetup{font=small}
    \caption{As in Fig. \ref{diff_spin}, but for a fixed spectral index value of $\Gamma=1.85$. The trend is similar, although the differences are now dampened.}
    \label{diff_spin_gamma}
\end{figure*}

\begin{figure*}[h!]
    \centering
    \begin{subfigure}[b]{0.45\textwidth}
        \centering
        \includegraphics[width=8.15cm]{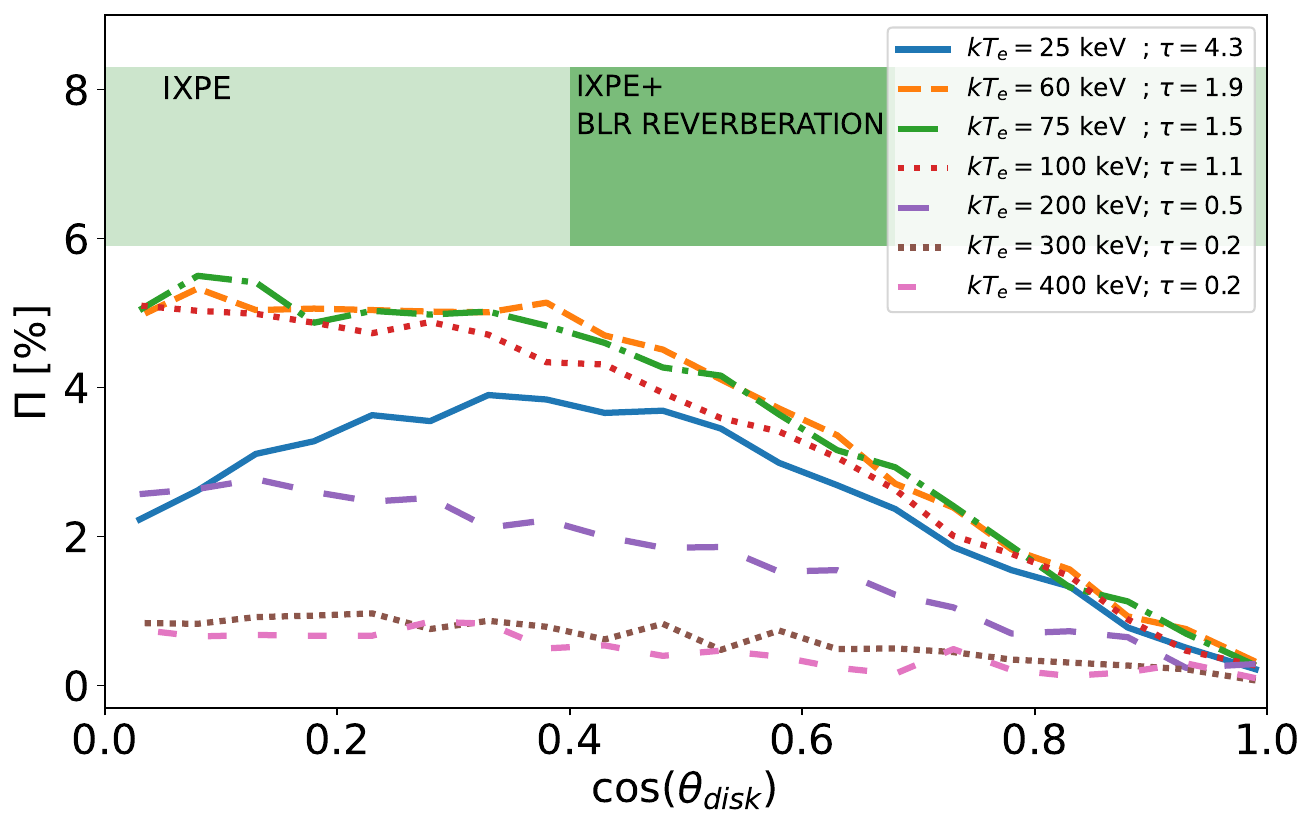} 
        \label{PD_sk_diff_kT_a0_gamma}
    \end{subfigure}
    \hfill
    \begin{subfigure}[b]{0.45\textwidth}
        \centering
        \includegraphics[width=8.15cm]{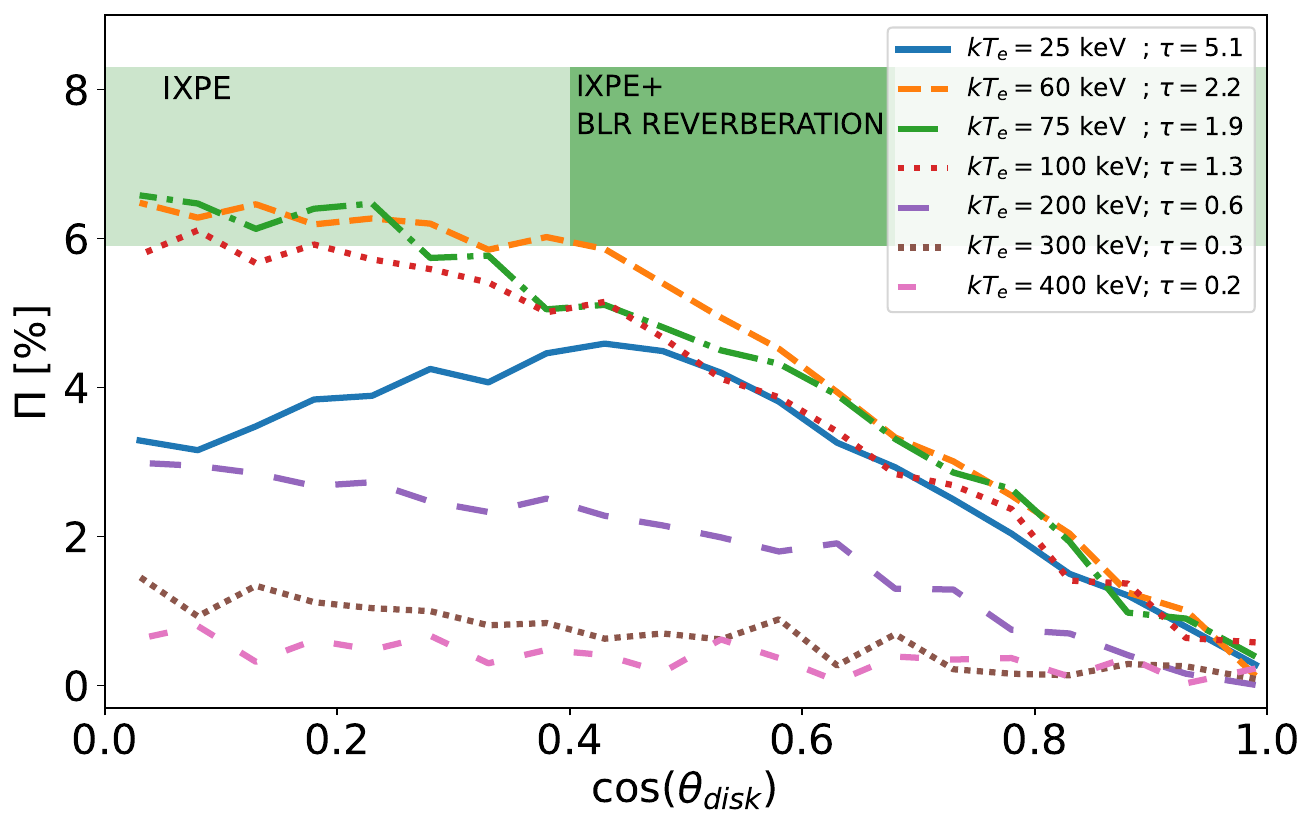} 
        \label{PD_sk_diff_kT_a998_gamma}
    \end{subfigure}
    \captionsetup{font=small}
    \caption{As in Fig. \ref{diff_kT}, but for a fixed spectral image value of $\Gamma=1.85$. Here, $\Pi$ increases up to$kT_e\sim75$ keV due to a lower $\tau$ needed to reproduce $\Gamma=1.85$, and then decreases because of the depolarizing effect of the Compton cross-section.}
    \label{diff_kT_gamma}
\end{figure*}

Next, we performed simulations using $\tau$ values that reproduce a primary continuum spectral index of $\Gamma$=$1.85\pm0.01$ for each different geometric and physical configuration, consistent with the best-fit value found for NGC~4151 (\citealt{Gianolli2023}). Initially, we fixed the electron temperature at 60 keV, as appropriate for this source, but we also explored a range of temperatures. We observed the same qualitative trends as in the previous cases, in which we left $\tau$ constant. For different opening angles (see Fig. \ref{diff_opang_gamma}), cases with higher $\tau$ than the baseline configuration result in lower $\Pi$, while cases in which $\tau$ is lower result in higher $\Pi$. Similar behavior is found when varying the inner disk radius (Fig. \ref{diff_disc_gamma}) and BH spin (Fig. \ref{diff_spin_gamma}). In contrast, a different behavior is obtained in the case of different coronal temperatures (Fig. \ref{diff_kT_gamma}): we observed an increase in $\Pi$ from $kT_{e}=25$ to $kT_{e}=75$ keV, followed by a decrease for higher temperatures. This is due to two opposing effects: for a given $\Gamma$, a higher temperature corresponds to a lower $\tau$ \citep{middei2019A&A...630A.131M}, but for higher temperature the depolarizing effect in the Compton cross-section becomes dominant.

Given the constraints on the polarization properties from the combined  IXPE analysis (\citealt{giano2024A&A...691A..29G}) and the disk inclination from broad-line region reverberation studies (\citealt{bentz2022ApJ...934..168B}) of NGC~4151, we were able to differentiate between models that reproduce the observed characteristics and those that do not. Moreover, all tested configurations result in $\Psi$ parallel to the disk axis, consistent
with IXPE data. We then restricted the allowed polarization fraction values to those indicated by \cite{Gianolli2023} (i.e.,  primary continuum polarization fraction $7.1\% \pm 1.2\%$), and the allowed source inclination to $47^{\circ}<\theta_{disk}<66^{\circ}$, as indicated by \cite{bentz2022ApJ...934..168B}. Both constraints are given at the $68\%$  confidence level and define the green regions shown in Fig. \ref{diff_opang_gamma}, \ref{diff_disc_gamma}, \ref{diff_spin_gamma} and \ref{diff_kT_gamma}. 

\begin{figure}[h!]
\includegraphics[width=8.15cm] {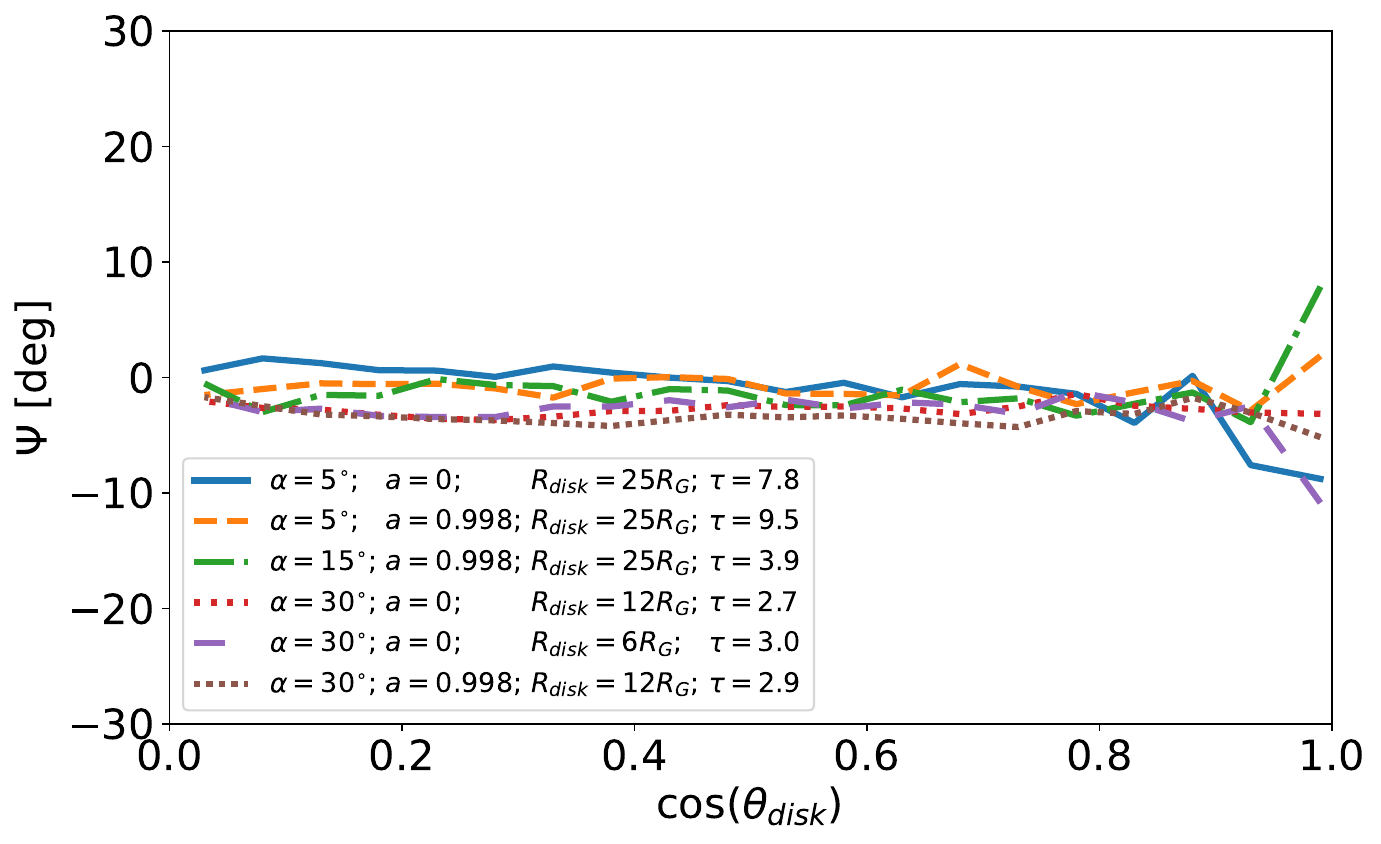}
\captionsetup{font=small}
\caption{Polarization angle ($\Psi$), summed between 2 and 8 keV, as a function of the inclination of the source, for models consistent with the constraints from IXPE analyses and BLR reverberation studies. For all models shown, the coronal temperature is set to $kT_{e}=60$ keV and the coronal inner radius is set to the ISCO radius corresponding to the BH spin ($R_{corona}=6 R_G$ when $a=0$ and $R_{corona}=1.24 R_G$ when $a=0.998$).}\label{comp}
\end{figure}

Comparing these constraints with the simulation results, we find that only a few configurations are able to reproduce the allowed $\Pi$ values. Generally, we find that low opening angles, high spins, and discs partially entering the corona better reproduce the observed polarization properties. Figure \ref{comp} shows $\Psi$ as a function of source inclination for those models that are able to reproduce the constraints on $\Pi$. Conversely, we can rule out a wedge corona model in the following scenarios: 
\begin{itemize}
    \item null spin combined with a high inner disk radius and opening angles greater than $5^{\circ}$ (left panel of Fig. \ref{diff_opang_gamma});
    \item null spin with $30^{\circ}$ opening angles for all tested electron temperatures (left panel of Fig. \ref{diff_kT_gamma});
    \item maximal spin with high disk inner radii and opening angles exceeding $15^{\circ}$ (right panel of Fig. \ref{diff_opang_gamma});
    \item maximal spin with high disk inner radii and $\alpha=30^{\circ}$ for every coronal temperature (right panel of Fig. \ref{diff_kT_gamma});
    \item maximal spin with inner disk radius reaching the ISCO and $\alpha=30^{\circ}$ (right panel of Fig. \ref{diff_disc_gamma}). 
\end{itemize}
Among the tested configurations, the one that best reproduces the allowed polarimetric values is characterized by $kT_e=60$ keV, $a=0.998$, $R_{\textnormal{disk}}=12$ $R_{\rm G}$, and $\alpha=30^{\circ}$ (see the right panels of Figs \ref{diff_disc_gamma} and \ref{diff_spin_gamma}). This particular configuration is compatible with the polarimetric constraints over a wider range of inclinations than any of the other scenarios considered. \\ It is worth noting that the best-fit configuration for NGC~4151, derived from IXPE, NuSTAR, and XMM-Newton data and producing a maximally spinning BH with $kT_e \sim 60$ keV (\citealt{giano2024A&A...691A..29G}), is also, within this work, one of the most compatible with the polarimetric and inclination constraints established for this source. We emphasize that this does not imply that the ultimate model has been found; rather, we have identified the most appropriate configuration from those evaluated here. Due to the extensive range of parameters, there may still be alternative physical and geometric configurations that can reproduce the observed data. Additionally, the constraints related to polarization and inclination are influenced by significant uncertainties. If we were to consider these at the $90\%$ and $99\%$ confidence levels instead of the $68\%$ currently considered, there would be more configurations capable of fitting the data and a broader range of allowed parameter space.

\section{Summary}
\label{discussion}
Using the Monte Carlo ray-tracing code \textsc{monk}, we investigated how the spectro-polarimetric properties of the AGN primary continuum in the $2-8$ keV energy band depend on the physical and geometric parameters of the wedge corona model. We find that both the primary continuum spectral index and the polarization degree vary with the coronal opening angle, temperature, optical depth, BH spin, and the AD inner radius. In contrast, in all tested cases, the polarization angle was parallel to the AD axis, as is generally found for radially extended coronal models. 

In particular, we explored:
\begin{itemize}
\item different values of the wedge opening angle ($5^{\circ}$, $15^{\circ}$, $30^{\circ}$, $45^{\circ}$, and $60^{\circ}$), observing a hardening of the $2-8$ keV spectrum and a decreasing polarization degree when moving to higher openings;
\item different disk inner radii (equal to the coronal external radius, going down to half the coronal depth and close to the ISCO). In this case, we found a softening of the spectrum and an increasing polarization degree for lower values of $R_{\textnormal{disk}}$;
\item different BH spin values (null and maximum), observing a softening of the spectrum and an increasing polarization fraction for highly rotating BHs;
\item  a range of coronal electrons temperatures (25, 60, 75, 100, 200, 300, and 400 keV). We report a hardening of the spectrum and a decrease of the polarization degree at higher temperatures;
\item various coronal optical depths (0.5, 1, 1.9, 2.5, 5, and 10). We also observed a hardening of the spectrum and a decrease of the polarization degree with increasing values of $\tau$.
\end{itemize}
Finally, we fine-tuned the coronal optical depth in order to reproduce the spectral index of NGC~4151, the only radio-quiet unobscured AGN with robust X-ray polarization detection by IXPE to date. Comparing the simulations with the IXPE results allowed us to exclude many configurations of the wedge model. We found that configurations with high spins and disks partially entering the corona better reproduce the observed polarization properties. The configuration that meets the polarimetric constraints across the widest range of inclinations is characterized by $kT_e=60$ keV, $a=0.998$, $R_{\textnormal{disk}}=12$ $R_{\rm G}$, and $\alpha=30^{\circ}$. This configuration aligns with the best-fit model found for NGC~4151 established through prior spectroscopic analysis. To help readers navigate the extensive parameter space and allow facile comparison between the presented figures, we summarize the full set of model parameters in Table \ref{summ_tab}.

\begin{acknowledgements}
The Imaging X-ray Polarimetry Explorer (IXPE) is a joint US and Italian mission. The US contribution is supported by the National Aeronautics and Space Administration (NASA) and led and managed by its Marshall Space Flight Center (MSFC), with industry partner Ball Aerospace (contract NNM15AA18C). DT, AG, SB, GM, and FU acknowledge financial support by the Italian Space Agency (Agenzia Spaziale Italiana, ASI) through the contract ASI-INAF-2022-19-HH.0. VEG acknowledges funding under NASA contract 80NSSC24K1403. WZ acknowledges NSFC grants 12333004, and support by the Strategic Priority Research Program of the Chinese Academy of Sciences, grant no. XDB0550200.
\end{acknowledgements}

\bibliographystyle{aa}
\bibliography{aa54138-25} 

\begin{appendix}
\section{Simulations parameters summary}
    \begin{table*}[t]
\captionsetup{font=small}
\caption{Summary table listing the full set of model parameters for each figure.}
\label{summ_tab}
\begin{tabular}{|cccccccc}
\hline
Figure                                                 & \multicolumn{1}{l}{$\alpha$ {[}deg{]}} & $a$                    & $R_\text{corona}$ $[R_\text{G}]$ & $R_\text{disk}$ $[R_\text{G}]$ & $kT_\text{e}$ {[}keV{]} & $\Gamma$ $(\tau=1.9)$ & \multicolumn{1}{c|}{$\tau$ $(\Gamma=1.85)$} \\ \hline \hline
\multirow{5}{*}{\ref{I_en_sk_a0_diff_ap_tau_const}, \ref{PA_opang}, \ref{diff_opang} (left panel), \ref{diff_opang_gamma} (left panel)} & 5                                      & \multirow{5}{*}{0}     & \multirow{5}{*}{6}               & \multirow{5}{*}{25}            & \multirow{5}{*}{60}         & 2.81                  & \multicolumn{1}{c|}{7.8}                    \\
                                                       & 15                                     &                        &                                  &                                &                             & 2.22                  & \multicolumn{1}{c|}{3.2}                    \\
                                                       & 30                                     &                        &                                  &                                &                             & 1.85                  & \multicolumn{1}{c|}{1.9}                    \\
                                                       & 45                                     &                        &                                  &                                &                             & 1.74                  & \multicolumn{1}{c|}{1.5}                    \\
                                                       & 60                                     &                        &                                  &                                &                             & 1.64                  & \multicolumn{1}{c|}{1.3}                    \\ \hline
\multirow{5}{*}{\ref{diff_opang} (right panel), \ref{diff_opang_gamma} (right panel)}     & 5                                      & \multirow{5}{*}{0.998} & \multirow{5}{*}{1.24}            & \multirow{5}{*}{25}            & \multirow{5}{*}{60}         & 2.85                  & \multicolumn{1}{c|}{9.5}                    \\
                                                       & 15                                     &                        &                                  &                                &                             & 2.27                  & \multicolumn{1}{c|}{3.9}                    \\
                                                       & 30                                     &                        &                                  &                                &                             & 1.97                  & \multicolumn{1}{c|}{2.2}                    \\
                                                       & 45                                     &                        &                                  &                                &                             & 1.81                  & \multicolumn{1}{c|}{1.7}                    \\
                                                       & 60                                     &                        &                                  &                                &                             & 1.72                  & \multicolumn{1}{c|}{1.5}                    \\ \hline
\multirow{3}{*}{\ref{diff_disc} (left panel), \ref{sca}, \ref{diff_disc_gamma} (left panel)}    & \multirow{3}{*}{30}                    & \multirow{3}{*}{0}     & \multirow{3}{*}{6}               & 25                             & \multirow{3}{*}{60}         & 1.85                  & \multicolumn{1}{c|}{1.9}                    \\
                                                       &                                        &                        &                                  & 12                             &                             & 2.11                  & \multicolumn{1}{c|}{2.7}                    \\
                                                       &                                        &                        &                                  & 6                              &                             & 2.16                  & \multicolumn{1}{c|}{3.0}                    \\ \hline
\multirow{3}{*}{\ref{diff_disc} (right panel), \ref{diff_disc_gamma} (right panel)}     & \multirow{3}{*}{30}                    & \multirow{3}{*}{0.998} & \multirow{3}{*}{1.24}            & 25                             & \multirow{3}{*}{60}         & 1.97                  & \multicolumn{1}{c|}{2.2}                    \\
                                                       &                                        &                        &                                  & 12                             &                             & 2.15                  & \multicolumn{1}{c|}{2.9}                    \\
                                                       &                                        &                        &                                  & 1.24                           &                             & 2.26                  & \multicolumn{1}{c|}{3.8}                    \\ \hline
\multirow{4}{*}{\ref{diff_spin} (left panel), \ref{diff_spin_gamma} (left panel)}       & 30                                     & \multirow{2}{*}{0}     & \multirow{2}{*}{6}               & \multirow{4}{*}{25}            & \multirow{4}{*}{60}         & 1.85                  & \multicolumn{1}{c|}{1.9}                    \\
                                                       & 45                                     &                        &                                  &                                &                             & 1.74                  & \multicolumn{1}{c|}{1.5}                    \\
                                                       & 30                                     & \multirow{2}{*}{0.998} & \multirow{2}{*}{1.24}            &                                &                             & 1.97                  & \multicolumn{1}{c|}{2.2}                    \\
                                                       & 45                                     &                        &                                  &                                &                             & 1.81                  & \multicolumn{1}{c|}{1.7}                    \\ \hline
\multirow{6}{*}{\ref{diff_spin} (right panel), \ref{diff_spin_gamma} (right panel)}     & \multirow{6}{*}{30}                    & \multirow{3}{*}{0}     & \multirow{3}{*}{6}               & 25                             & \multirow{6}{*}{60}         & 1.85                  & \multicolumn{1}{c|}{1.9}                    \\
                                                       &                                        &                        &                                  & 12                             &                             & 2.11                  & \multicolumn{1}{c|}{2.7}                    \\
                                                       &                                        &                        &                                  & 6                              &                             & 2.16                  & \multicolumn{1}{c|}{3.0}                    \\
                                                       &                                        & \multirow{3}{*}{0.998} & \multirow{3}{*}{1.24}            & 25                             &                             & 1.97                  & \multicolumn{1}{c|}{2.2}                    \\
                                                       &                                        &                        &                                  & 12                             &                             & 2.15                  & \multicolumn{1}{c|}{2.9}                    \\
                                                       &                                        &                        &                                  & 1.24                           &                             & 2.26                  & \multicolumn{1}{c|}{3.8}                    \\ \hline
\multirow{7}{*}{\ref{diff_kT} (left panel), \ref{diff_kT_gamma} (left panel)}       & \multirow{7}{*}{30}                    & \multirow{7}{*}{0}     & \multirow{7}{*}{6}               & \multirow{7}{*}{25}            & 25                          & 2.66                  & \multicolumn{1}{c|}{4.3}                    \\
                                                       &                                        &                        &                                  &                                & 60                          & 1.85                  & \multicolumn{1}{c|}{1.9}                    \\
                                                       &                                        &                        &                                  &                                & 75                          & 1.77                  & \multicolumn{1}{c|}{1.5}                    \\
                                                       &                                        &                        &                                  &                                & 100                         & 1.63                  & \multicolumn{1}{c|}{1.1}                    \\
                                                       &                                        &                        &                                  &                                & 200                         & 1.43                  & \multicolumn{1}{c|}{0.5}                    \\
                                                       &                                        &                        &                                  &                                & 300                         & 1.32                  & \multicolumn{1}{c|}{0.2}                    \\
                                                       &                                        &                        &                                  &                                & 400                         & 1.24                  & \multicolumn{1}{c|}{0.2}                    \\ \hline
\multirow{7}{*}{\ref{diff_kT} (right panel), \ref{diff_kT_gamma} (right panel)}     & \multirow{7}{*}{30}                    & \multirow{7}{*}{0.998} & \multirow{7}{*}{1.24}            & \multirow{7}{*}{25}            & 25                          & 2.64                  & \multicolumn{1}{c|}{5.1}                    \\
                                                       &                                        &                        &                                  &                                & 60                          & 1.97                  & \multicolumn{1}{c|}{2.2}                    \\
                                                       &                                        &                        &                                  &                                & 75                          & 1.84                  & \multicolumn{1}{c|}{1.9}                    \\
                                                       &                                        &                        &                                  &                                & 100                         & 1.70                  & \multicolumn{1}{c|}{1.3}                    \\
                                                       &                                        &                        &                                  &                                & 200                         & 1.47                  & \multicolumn{1}{c|}{0.6}                    \\
                                                       &                                        &                        &                                  &                                & 300                         & 1.37                  & \multicolumn{1}{c|}{0.3}                    \\
                                                       &                                        &                        &                                  &                                & 400                         & 1.28                  & \multicolumn{1}{c|}{0.2}                    \\ \hline
\multirow{6}{*}{\ref{comp}}                                    & 5                                      & \multirow{3}{*}{0}     & \multirow{3}{*}{6}               & 25                             & \multirow{6}{*}{60}         & \multirow{6}{*}{/}    & \multicolumn{1}{c|}{7.8}                    \\
                                                       & 30                                     &                        &                                  & 12                             &                             &                       & \multicolumn{1}{c|}{2.7}                    \\
                                                       & 30                                     &                        &                                  & 6                              &                             &                       & \multicolumn{1}{c|}{3.0}                    \\
                                                       & 5                                      & \multirow{3}{*}{0.998} & \multirow{3}{*}{1.24}            & 25                             &                             &                       & \multicolumn{1}{c|}{9.5}                    \\
                                                       & 15                                     &                        &                                  & 25                             &                             &                       & \multicolumn{1}{c|}{3.9}                    \\
                                                       & 30                                     &                        &                                  & 12                             &                             &                       & \multicolumn{1}{c|}{2.9}                    \\ \hline
\multicolumn{8}{c}{}                                                                                                                                                                                                                                                                            \\ \hline
Figure                                                 & \multicolumn{1}{l}{$\alpha$ {[}deg{]}} & $a$                    & $R_\text{corona}$ $[R_\text{G}]$ & $R_\text{disk}$ $[R_\text{G}]$ & $kT_\text{e}$ {[}keV{]} & ${\Gamma}$            & \multicolumn{1}{c|}{${\tau}$}               \\ \hline \hline
\multirow{6}{*}{\ref{diff_tau}}                                     & \multirow{6}{*}{30}                    & \multirow{6}{*}{0}     & \multirow{6}{*}{6}               & \multirow{6}{*}{25}            & \multirow{6}{*}{60}         & 2.79                  & \multicolumn{1}{c|}{0.5}                    \\
                                                       &                                        &                        &                                  &                                &                             & 1.94                  & \multicolumn{1}{c|}{1.0}                    \\
                                                       &                                        &                        &                                  &                                &                             & 1.85                  & \multicolumn{1}{c|}{1.9}                    \\
                                                       &                                        &                        &                                  &                                &                             & 1.63                  & \multicolumn{1}{c|}{2.5}                    \\
                                                       &                                        &                        &                                  &                                &                             & 1.37                  & \multicolumn{1}{c|}{5.0}                    \\
                                                       &                                        &                        &                                  &                                &                             & 1.22                  & \multicolumn{1}{c|}{10.0}                   \\ \hline
\end{tabular}
\tablefoot{Except the last row (referring to Fig. \ref{diff_tau}) the last column refers to the $\tau$ values reproducing $\Gamma=1.85$ (plots showing the polarization properties assuming NGC~4151 physical parameters) and the second to last column to the $\Gamma$ values obtained keeping $\tau=1.9$.}
\end{table*} 
\end{appendix}

\end{document}